\documentclass[%
reprint,
superscriptaddress,
amsmath,amssymb,
aps,
]{revtex4-2}

\usepackage{graphicx}% Include figure files
\usepackage{dcolumn}% Align table columns on decimal point
\usepackage{bm}% bold math
\usepackage{multirow}
\usepackage{float}
\usepackage{CJK}
\usepackage{array}
\begin{document}

\preprint{APS/123-QED}

\title{Search for Neutrinoless Double-Beta Decay of 
   $^{76}$Ge with a Natural Broad Energy Germanium Detector}% Force line breaks with \\

\author{W. H. Dai}
\author{H. Ma}
 \email{mahao@tsinghua.edu.cn}
\author{Q. Yue}
 \email{yueq@mail.tsinghua.edu.cn}
\author{Z. She}
\author{K. J. Kang}
\author{Y. J. Li}
\affiliation{Key Laboratory of Particle and Radiation Imaging 
(Ministry of Education) and Department of Engineering Physics, 
Tsinghua University, Beijing 100084}

\author{M. Agartioglu}
\altaffiliation[Participating as a member of ]{TEXONO Collaboration}
\affiliation{Institute of Physics, Academia Sinica, Taipei 11529}

\author{H. P. An}
\affiliation{Key Laboratory of Particle and Radiation Imaging 
(Ministry of Education) and Department of Engineering Physics, 
Tsinghua University, Beijing 100084}
\affiliation{Department of Physics, Tsinghua University, Beijing 100084}

\author{J. P. Chang}
\affiliation{NUCTECH Company, Beijing 100084}

\author{Y. H. Chen}
\affiliation{YaLong River Hydropower Development Company, Chengdu 610051}

\author{J. P. Cheng}
\affiliation{Key Laboratory of Particle and Radiation Imaging 
(Ministry of Education) and Department of Engineering Physics, 
Tsinghua University, Beijing 100084}
\affiliation{College of Nuclear Science and Technology, Beijing Normal University, Beijing 100875}

\author{Z. Deng}
\affiliation{Key Laboratory of Particle and Radiation Imaging 
(Ministry of Education) and Department of Engineering Physics, 
Tsinghua University, Beijing 100084}

\author{C. H. Fang}
\affiliation{College of Physics, Sichuan University, Chengdu 610065}

\author{X. P. Geng}
\affiliation{Key Laboratory of Particle and Radiation Imaging 
(Ministry of Education) and Department of Engineering Physics, 
Tsinghua University, Beijing 100084}

\author{H. Gong}
\affiliation{Key Laboratory of Particle and Radiation Imaging 
(Ministry of Education) and Department of Engineering Physics, 
Tsinghua University, Beijing 100084}

\author{Q. J. Guo}
\affiliation{School of Physics, Peking University, Beijing 100871}

% \author{X. P. Geng}
% \affiliation{Key Laboratory of Particle and Radiation Imaging 
% (Ministry of Education) and Department of Engineering Physics, 
% Tsinghua University, Beijing 100084}

\author{X. Y. Guo}
\affiliation{YaLong River Hydropower Development Company, Chengdu 610051}

\author{L. He}
\affiliation{NUCTECH Company, Beijing 100084}

\author{S. M. He}
\affiliation{YaLong River Hydropower Development Company, Chengdu 610051}

\author{J. W. Hu}
\affiliation{Key Laboratory of Particle and Radiation Imaging 
(Ministry of Education) and Department of Engineering Physics, 
Tsinghua University, Beijing 100084}

\author{H. X. Huang}
\affiliation{Department of Nuclear Physics, China Institute of Atomic Energy, Beijing 102413}

\author{T. C. Huang}
\affiliation{Sino-French Institute of Nuclear and Technology, Sun Yat-sen University, Zhuhai 519082}

\author{H. T. Jia}
\author{X. Jiang}
\affiliation{College of Physics, Sichuan University, Chengdu 610065}

\author{H. B. Li}
\altaffiliation[Participating as a member of ]{TEXONO Collaboration}
\affiliation{Institute of Physics, Academia Sinica, Taipei 11529}

% \author{H. Li}
% \affiliation{NUCTECH Company, Beijing 100084}

\author{J. M. Li}
\author{J. Li}
\affiliation{Key Laboratory of Particle and Radiation Imaging 
(Ministry of Education) and Department of Engineering Physics, 
Tsinghua University, Beijing 100084}

\author{Q. Y. Li}
\author{R. M. J. Li}
\affiliation{College of Physics, Sichuan University, Chengdu 610065}

% \author{X. Li}
% \affiliation{Department of Nuclear Physics, China Institute of Atomic Energy, Beijing 102413}

\author{X. Q. Li}
\affiliation{School of Physics, Nankai University, Tianjin 300071}

\author{Y. L. Li}
\affiliation{Key Laboratory of Particle and Radiation Imaging 
(Ministry of Education) and Department of Engineering Physics, 
Tsinghua University, Beijing 100084}

\author{Y. F. Liang}
\affiliation{Key Laboratory of Particle and Radiation Imaging 
(Ministry of Education) and Department of Engineering Physics, 
Tsinghua University, Beijing 100084}

\author{B. Liao}
\affiliation{College of Nuclear Science and Technology, Beijing Normal University, Beijing 100875}

\author{F. K. Lin}
\altaffiliation[Participating as a member of ]{TEXONO Collaboration}
\affiliation{Institute of Physics, Academia Sinica, Taipei 11529}

\author{S. T. Lin}
\author{S. K. Liu}
\affiliation{College of Physics, Sichuan University, Chengdu 610065}

\author{Y. D. Liu}
\affiliation{College of Nuclear Science and Technology, Beijing Normal University, Beijing 100875}

\author{Y. Liu}
\affiliation{College of Physics, Sichuan University, Chengdu 610065}

\author{Y. Y. Liu}
\affiliation{College of Nuclear Science and Technology, Beijing Normal University, Beijing 100875}

\author{Z. Z. Liu}
\affiliation{Key Laboratory of Particle and Radiation Imaging 
(Ministry of Education) and Department of Engineering Physics, 
Tsinghua University, Beijing 100084}

\author{Y. C. Mao}
\affiliation{School of Physics, Peking University, Beijing 100871}

\author{Q. Y. Nie}
\affiliation{Key Laboratory of Particle and Radiation Imaging 
(Ministry of Education) and Department of Engineering Physics, 
Tsinghua University, Beijing 100084}

\author{J. H. Ning}
\affiliation{YaLong River Hydropower Development Company, Chengdu 610051}

\author{H. Pan}
\affiliation{NUCTECH Company, Beijing 100084}

\author{N. C. Qi}
\affiliation{YaLong River Hydropower Development Company, Chengdu 610051}

\author{J. Ren}
\author{X. C. Ruan}
\affiliation{Department of Nuclear Physics, China Institute of Atomic Energy, Beijing 102413}

\author{K. Saraswat}
\altaffiliation[Participating as a member of ]{TEXONO Collaboration}
\affiliation{Institute of Physics, Academia Sinica, Taipei 11529}

\author{V. Sharma}
\altaffiliation[Participating as a member of ]{TEXONO Collaboration}
\affiliation{Institute of Physics, Academia Sinica, Taipei 11529}
\affiliation{Department of Physics, Banaras Hindu University, Varanasi 221005}

\author{M. K. Singh}
\altaffiliation[Participating as a member of ]{TEXONO Collaboration}
\affiliation{Institute of Physics, Academia Sinica, Taipei 11529}
\affiliation{Department of Physics, Banaras Hindu University, Varanasi 221005}

\author{T. X. Sun}
\affiliation{College of Nuclear Science and Technology, Beijing Normal University, Beijing 100875}

\author{C. J. Tang}
\affiliation{College of Physics, Sichuan University, Chengdu 610065}

\author{W. Y. Tang}
\author{Y. Tian}
\affiliation{Key Laboratory of Particle and Radiation Imaging 
(Ministry of Education) and Department of Engineering Physics, 
Tsinghua University, Beijing 100084}

\author{G. F. Wang}
\affiliation{College of Nuclear Science and Technology, Beijing Normal University, Beijing 100875}

\author{L. Wang}
\affiliation{Department of Physics, Beijing Normal University, Beijing 100875}

\author{Q. Wang}
\author{Y. Wang}
\affiliation{Key Laboratory of Particle and Radiation Imaging 
(Ministry of Education) and Department of Engineering Physics, 
Tsinghua University, Beijing 100084}
\affiliation{Department of Physics, Tsinghua University, Beijing 100084}

\author{Y. X. Wang}
\affiliation{School of Physics, Peking University, Beijing 100871}

\author{H. T. Wong}
\altaffiliation[Participating as a member of ]{TEXONO Collaboration}
\affiliation{Institute of Physics, Academia Sinica, Taipei 11529}

\author{S. Y. Wu}
\affiliation{YaLong River Hydropower Development Company, Chengdu 610051}

\author{Y. C. Wu}
\affiliation{Key Laboratory of Particle and Radiation Imaging 
(Ministry of Education) and Department of Engineering Physics, 
Tsinghua University, Beijing 100084}

\author{H. Y. Xing}
\affiliation{College of Physics, Sichuan University, Chengdu 610065}

\author{R. Xu}
\affiliation{Key Laboratory of Particle and Radiation Imaging 
(Ministry of Education) and Department of Engineering Physics, 
Tsinghua University, Beijing 100084}

\author{Y. Xu}
\affiliation{School of Physics, Nankai University, Tianjin 300071}

\author{T. Xue}
\affiliation{Key Laboratory of Particle and Radiation Imaging 
(Ministry of Education) and Department of Engineering Physics, 
Tsinghua University, Beijing 100084}

\author{Y. L. Yan}
\affiliation{College of Physics, Sichuan University, Chengdu 610065}

\author{L. T. Yang}
\affiliation{Key Laboratory of Particle and Radiation Imaging 
(Ministry of Education) and Department of Engineering Physics, 
Tsinghua University, Beijing 100084}

\author{C. H. Yeh}
\altaffiliation[Participating as a member of ]{TEXONO Collaboration}
\affiliation{Institute of Physics, Academia Sinica, Taipei 11529}

\author{N. Yi}
\affiliation{Key Laboratory of Particle and Radiation Imaging 
(Ministry of Education) and Department of Engineering Physics, 
Tsinghua University, Beijing 100084}

\author{C. X. Yu}
\affiliation{School of Physics, Nankai University, Tianjin 300071}

\author{H. J. Yu}
\affiliation{NUCTECH Company, Beijing 100084}

\author{J. F. Yue}
\affiliation{YaLong River Hydropower Development Company, Chengdu 610051}

\author{M. Zeng}
\author{Z. Zeng}
\author{B. T. Zhang}
\affiliation{Key Laboratory of Particle and Radiation Imaging 
(Ministry of Education) and Department of Engineering Physics, 
Tsinghua University, Beijing 100084}

\author{F. S. Zhang}
\affiliation{College of Nuclear Science and Technology, Beijing Normal University, Beijing 100875}

\author{L. Zhang}
\affiliation{College of Physics, Sichuan University, Chengdu 610065}

\author{Z. H. Zhang}
\author{Z. Y. Zhang}
\affiliation{Key Laboratory of Particle and Radiation Imaging 
(Ministry of Education) and Department of Engineering Physics, 
Tsinghua University, Beijing 100084}

\author{K. K. Zhao}
\affiliation{College of Physics, Sichuan University, Chengdu 610065}

\author{M. G. Zhao}
\affiliation{School of Physics, Nankai University, Tianjin 300071}

\author{J. F. Zhou}
\affiliation{YaLong River Hydropower Development Company, Chengdu 610051}

\author{Z. Y. Zhou}
\affiliation{Department of Nuclear Physics, China Institute of Atomic Energy, Beijing 102413}

\author{J. J. Zhu}
\affiliation{College of Physics, Sichuan University, Chengdu 610065}

\collaboration{CDEX Collaboration}%\noaffiliation

\date{\today}% It is always \today, today,

\begin{abstract}
% A natural broad energy germanium (BEGe) detector is operated to search for the neutrinoless double-beta
% (0\textup{$\nu\beta\beta$}) decay of $^{76}$Ge in the China Jinping Underground Laboratory (CJPL).
A natural broad energy germanium (BEGe) detector is operated 
in the China Jinping Underground Laboratory (CJPL)
for a feasibility study of building the next generation 
experiment of 
the neutrinoless double-beta
(0\textup{$\nu\beta\beta$}) decay of $^{76}$Ge.
% The primary goal of this run is to show the feasibility of 
% building a next generation 0\textup{$\nu\beta\beta$} experiment 
% and test the relevant technologies.
The setup of the prototype facility, characteristics of the BEGe detector, 
background reduction methods and data analysis are described in this paper.
A background index of 6.4$\times$10$^{-3}$ counts/(keV$\cdot$kg$\cdot$day)
is achieved and 1.86 times lower than our previous result of the CDEX-1 detector.
No signal is observed with an exposure of 186.4 kg$\cdot$day, 
thus a limit on the half life of $^{76}$Ge 0\textup{$\nu\beta\beta$} decay 
is set at T$_{1/2}^{0\nu}$ $>$ 5.62×10$^{22}$ yr at 90\% C.L..
The limit corresponds to an effective Majorana neutrino mass 
in the range of 4.6 $\sim$ 10.3 eV, dependent on the nuclear matrix elements.
\begin{description}
\item[Key words]
Neutrinoless double-beta decay, BEGe, $^{76}$Ge, CJPL. 
\end{description}
\end{abstract}

% \keywords{showkeys}%Use showkeys class option if keyword
%                               % display desired
\maketitle

%\tableofcontents

% \section{\label{sec:level1}First-level heading:\protect\\ The line
% break was forced \lowercase{via} \textbackslash\textbackslash}
\section{\label{sec:1}Introduction}

Evidences for non-zero mass neutrinos have been provided by 
the atmospheric and solar neutrino oscillation experiments
[\onlinecite{bib:1}-\onlinecite{bib:4}]
over the last two decades. Neutrinos can obtain their masses by a 
Majorana mass term if they are their own anti-particles
[\onlinecite{bib:5}]. The Majorana nature of the neutrinos leads to lepton number violation 
and naturally emerges in many beyond-the-Standard Model theories
[\onlinecite{bib:6}].  It also emerges in leading theories that 
explain the dominance of matter over antimatter in the Universe
[\onlinecite{bib:7},\onlinecite{bib:8}].

The search for neutrinoless double-beta (0\textup{$\nu\beta\beta$}) 
decay is considered the most promising way to prove the Majorana nature of neutrinos
[\onlinecite{bib:9}]. Furthermore, a measurement of the 0\textup{$\nu\beta\beta$} decay rate, 
which depends on the effective Majorana mass, can indicate the mass hierarchy 
and the absolute mass scale of neutrinos.

Assuming the Majorana nature of neutrinos, the neutrinoless double-beta decay,
$(A,Z) \to (A,Z+2)+2e^{-}$, is permitted in 2\textup{$\nu\beta\beta$} decay isotopes
[\onlinecite{bib:10}]. 
% A number of experiments have conducted 0\textup{$\nu\beta\beta$} decay 
% search in various candidate isotopes, for instance
A huge experimental effort is ongoing to search for 
0\textup{$\nu\beta\beta$} decay in various 
candidate isotopes, for instance
$^{76}$Ge [\onlinecite{bib:11}-\onlinecite{bib:14}], 
$^{136}$Xe [\onlinecite{bib:15},\onlinecite{bib:16}], 
$^{130}$Te [\onlinecite{bib:17},\onlinecite{bib:18}], 
$^{100}$Mo [\onlinecite{bib:19},\onlinecite{bib:20}],
via different detection technologies, including 
semiconductor detector [\onlinecite{bib:11}-\onlinecite{bib:13}], 
time projection chamber [\onlinecite{bib:15}] 
and cryogenic bolometer [\onlinecite{bib:17},\onlinecite{bib:20},\onlinecite{bib:21}].

High purity germanium (HPGe), serving as both target nuclei and detector,
is an ideal medium for detecting 0\textup{$\nu\beta\beta$} decays
because of its high energy resolution, low internal background, 
and high detection efficiency [\onlinecite{bib:22}].
Several experiments have been searching for 
0$\nu\beta\beta$ decay
in $^{76}$Ge via the HPGe technology, such as
G\textsc{erda} [\onlinecite{bib:12}] and 
% MAJORANA DEMONSTRATOR [\onlinecite{bib:13}].
M\textsc{ajorana} D\textsc{emonstrator} [\onlinecite{bib:13}].
Currently, the G\textsc{erda} experiment, 
operating enriched germanium detector array in liquid argon to detect 
0\textup{$\nu\beta\beta$} decay of $^{76}$Ge, 
achieves the lowest background level in the
0\textup{$\nu\beta\beta$} decay Q value (Q$_{\beta\beta}$=2039 keV)
energy region and gives the 
most stringent constraint on the 
$^{76}$Ge 0\textup{$\nu\beta\beta$} half-life
(T$_{1/2}^{0\nu}$ $>$ 1.8$\times$10$^{26}$yr) [\onlinecite{bib:12}].
The G\textsc{erda} and the Majorana collaborations are now merged into the Legend collaboration
and are proposing a 200 kg-scale 0\textup{$\nu\beta\beta$} experiment (Legend-200)
aiming at setting the 0\textup{$\nu\beta\beta$} decay half-life limit of 
$^{76}$Ge at 10$^{27}$ yr [\onlinecite{bib:14}].

The CDEX collaboration has given its first 0\textup{$\nu\beta\beta$} limit of 
T$_{1/2}^{0\nu}$ $>$ 6.4$\times$10$^{22}$ yr for a p-type point contact high-purity 
germanium detector [\onlinecite{bib:23}]. 
A next-generation 0$\nu\beta\beta$ experiment CDEX-300$\nu$ has been proposed 
in CJPL-II [\onlinecite{bib:24}]. 
The CDEX-300$\nu$ experiment aims at achieving a discovery potential that 
reaches the inverted-ordering neutrino mass scale region with 1-ton$\cdot$yr exposure.

In this work, we set up a prototype facility in CJPL to study 
characteristics of a BEGe detector and novel background suppression techniques for future 
applications in the CDEX-300$\nu$ experiment. 
Data acquisition, analysis and pulse shape discrimination procedures are established and tested. 
And a 0$\nu\beta\beta$ result is given by analyzing a 186.4 kg$\cdot$day exposure data using 
an unbinned extended profile likelihood method.

\section{\label{sec:2}Experiment Setup}

\begin{figure}[!htb]
\includegraphics
[width=0.8\hsize]
{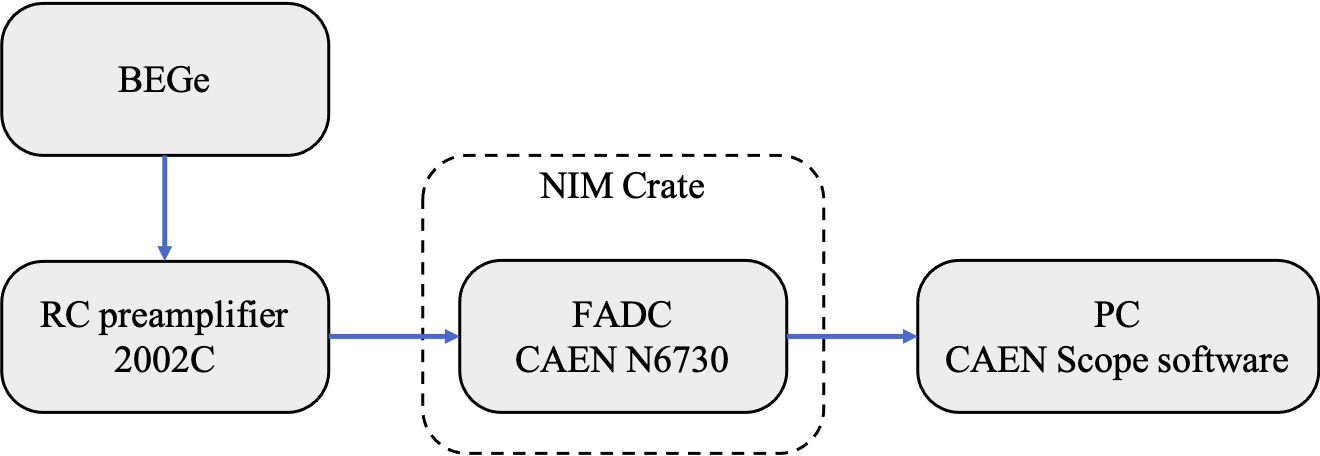}% Here is how to import EPS art
\caption{\label{fig:DAQ} Schematic diagram of the DAQ system.}
\end{figure}

A 1088.5 g natural low background broad energy germanium (BEGe) detector 
made by CANBERRA is used in our experiment. 
It is fabricated with a natural p-type germanium crystal 
with 91.1 mm in diameter and 31.4 mm in height.
The atom fraction of $^{76}$Ge in the crystal is 7.83\%. 
The BEGe detector operates at 4500 V high voltage.
The output from the p+ electrode is fed into a Canberra 2002C RC preamplifier 
to cover a wide dynamic energy range of up to 3.5 MeV 
for the 0$\nu\beta\beta$ decay search experiment.
The preamplifier output is digitized by a flash analog-to-digital convertor (FADC) 
at a 500 MHz sampling rate and recorded by the CAEN Scope software.
The trigger threshold of FADC is set to only record events with energy above 500 keV, 
and the trigger rate is approximately 0.005 cps during data taking. 
A schematic diagram of the data acquisition (DAQ) system 
is shown in Fig.~\ref{fig:DAQ}.

An experiment setup is built in the polyethylene (PE) room of the CJPL-I experiment hall. 
The over 2400 m rock overburden provides natural shields against the cosmic rays, 
and the cosmic muon flux in CJPL is about 
(2.0$\pm$0.4)$\times$10$^{-10}$ cm$^{-2}$s$^{-1}$ [\onlinecite{bib:25}].
Environmental neutrons are shielded by the 1m thick wall of the PE room, 
the thermal neutron flux inside the PE room is measured to be 
(3.18$\pm$0.97)$\times$10$^{-8}$ cm$^{-2}$s$^{-1}$ [\onlinecite{bib:26}].

\begin{figure}[!htb]
\includegraphics
[width=0.8\hsize]
{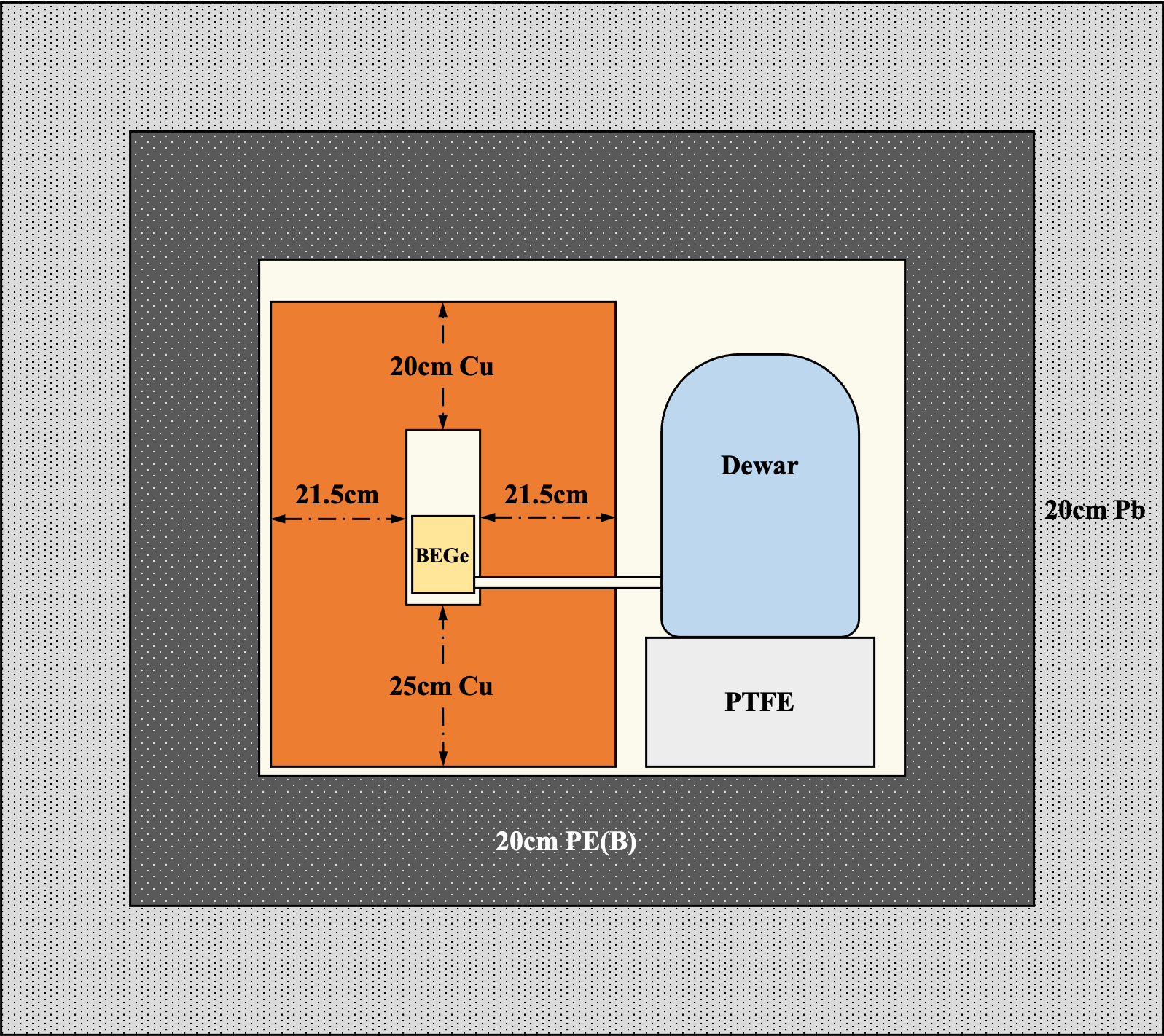}
\caption{\label{fig:setup} Schematic diagram of the experimental set up; 
The setup is located in a PE room (not shown) with 1 m thick PE wall.}
\end{figure}

A passive shielding structure is built to shield the ambient radioactivity. 
As shown in Fig.~\ref{fig:setup}, the detector crystal is shielded with a 20 cm lead, 
a 20 cm borated polyethylene, and a minimum of 20 cm copper from outside to inside.
The outmost 20 cm lead is used to shield the ambient gamma rays. 
The middle 20 cm borated polyethylene acts as a thermal neutron absorber. 
The innermost copper shield is made of low background oxygen-free high conductivity (OFHC) copper 
to shield the residual gamma rays surviving the outer shields. 
The space within the copper shields is continuously flushed with high purity nitrogen gas 
from a pressurized Dewar to exclude the radon.

\section{\label{sec:3}Data Analysis}

\subsection{\label{sec:3.1}Event selection and energy calibration}

\begin{figure}[htbp]
\includegraphics
[width=1.0\hsize]
{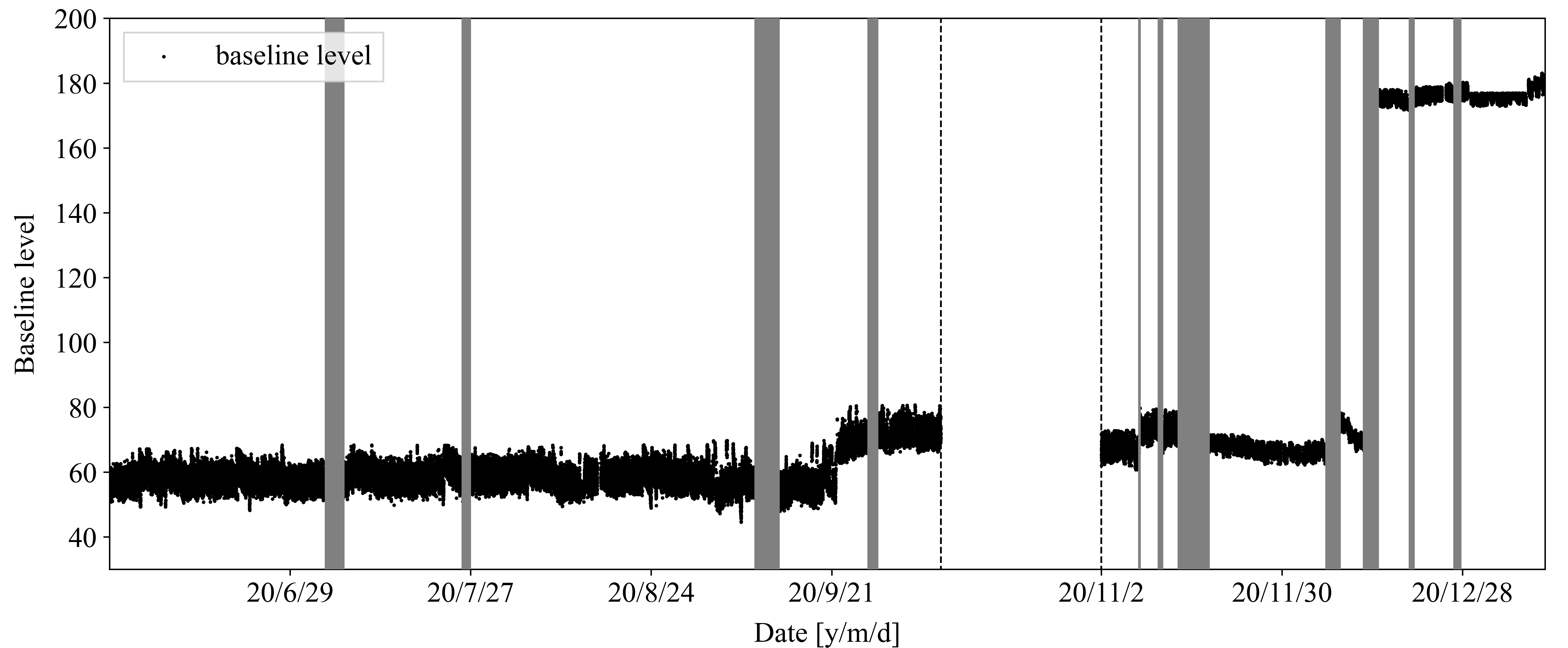}
\caption{\label{fig:baseline} Baseline level of the detector during data taking. 
The shadow regions are excluded from analysis because of their unstable baseline level. 
A major shift of baseline level on 2020/12/05 is caused by an accidental power failure.}
\end{figure}

The baseline level of the 500 MHz FADC is used to monitor the working condition of the detector, 
as shown in Fig.~\ref{fig:baseline}. 
Data taking starts on 2020/06/01 and ends on 2021/01/10. 
The gap from 2020/10/08 to 2020/11/02 was due to 
the unstable power supply caused by the construction of CJPL-II. 
Periods with unstable baseline levels are excluded from analysis 
(shadow regions in Fig.~\ref{fig:baseline}). 
On December 15, 2020, an accidental power failure caused a significant shift in the baseline level.

After excluding the shadow region in Fig.~\ref{fig:baseline}, 
the remaining 186.4 kg$\cdot$day exposure data is divided into 9 datasets 
depending on the time and the baseline level of the detector. 
Data selections and the energy calibration are performed independently in each dataset.

\begin{figure}[htbp]
\includegraphics
[width=1.0\hsize]
{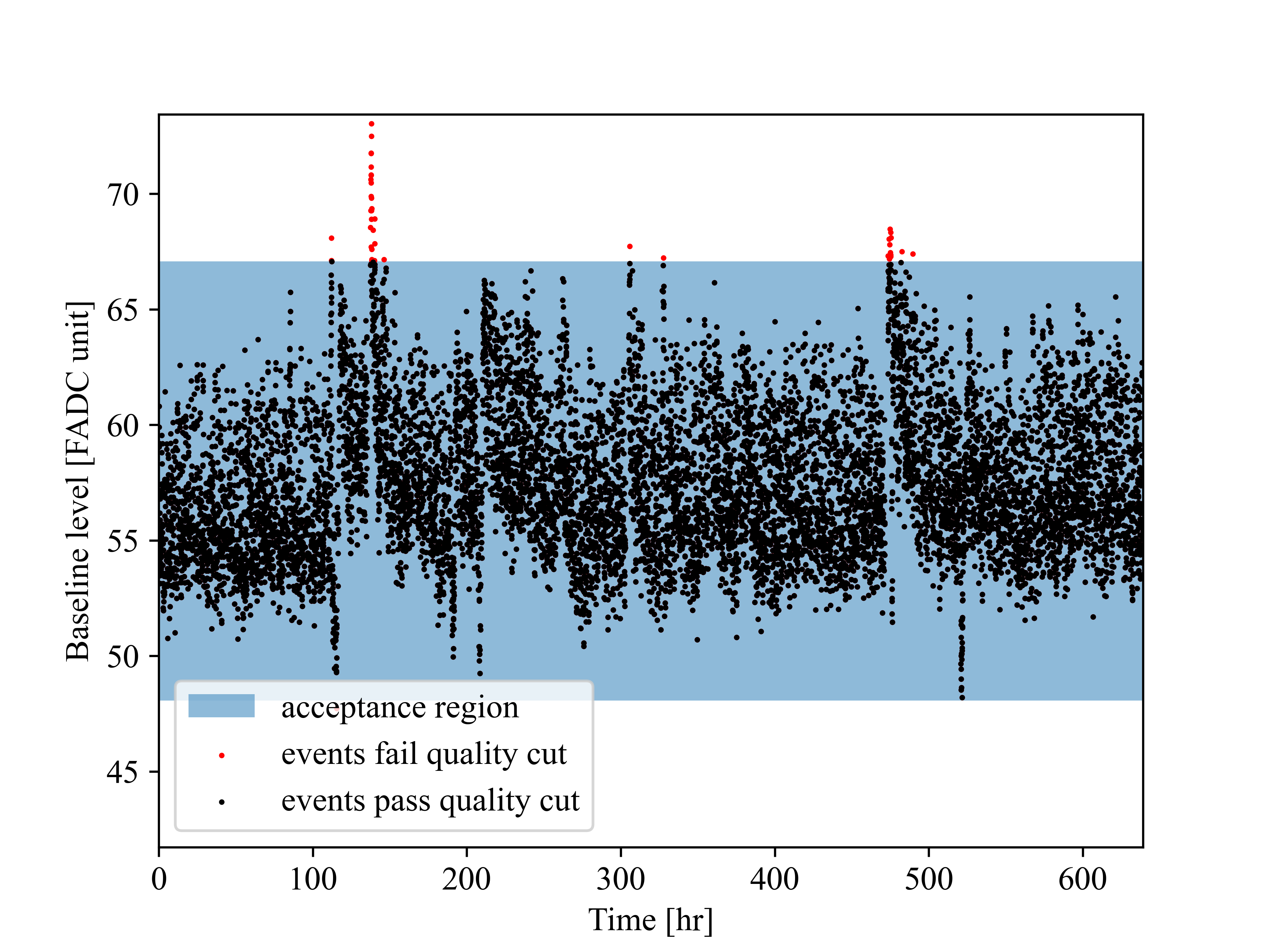}
\caption{\label{fig:QC} Baseline levels and the data quality cut of one dataset, 
red points are events that fail the adta quality cuts, 
the acceptance region is labeled in blue.}
\end{figure}

The recorded events are selected by a noise cut and a data quality cut 
to remove noise events and events with abnormal baseline levels.  
Unphysical events are almost noise bursts with 
minimum signal values much lower than the baseline, 
while the physical events have minimum signal values 
around the baseline. Therefore, unphysical events can 
be rejected by the noise cut:
events with minimum pulse values much lower than their baseline levels (10\% of trigger threshold) 
are rejected. Events with baseline level not in $\pm$3 times standard deviation of 
the average baseline level are rejected by the data quality cut. 
Fig.~\ref{fig:QC} shows baseline levels and the acceptance region of the data quality cut for one dataset.

\begin{figure}[!htb]
\includegraphics
[width=1.0\hsize]
{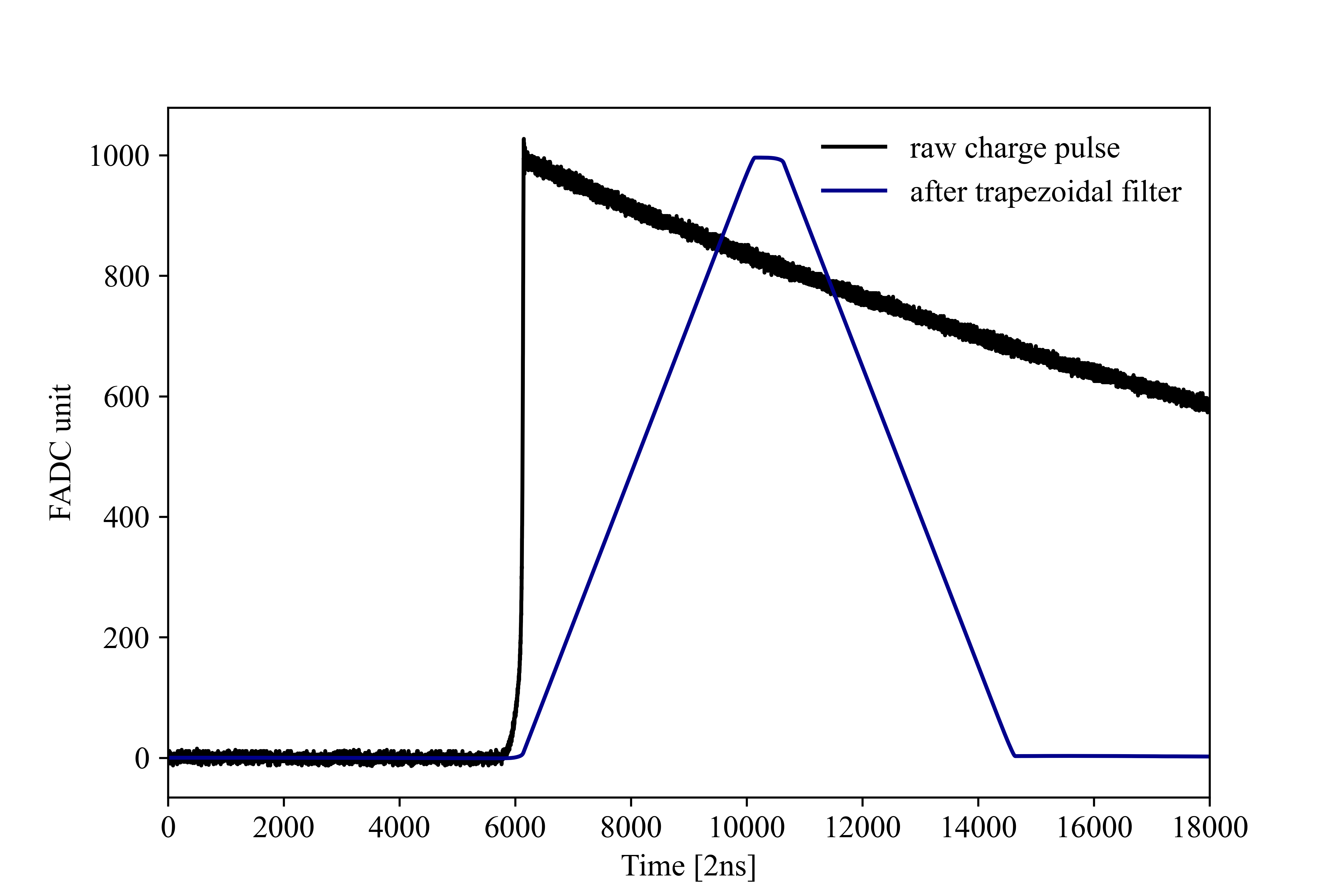}
\caption{\label{fig:Filter} An example of a charge pulse before/after 
the trapezoidal filter, the baseline of the charge pulse has been subtracted.}
\end{figure}

Amplitudes are extracted from the remaining charge pulses via a 
trapezoidal filter [\onlinecite{bib:27},\onlinecite{bib:28}]. 
The filter parameters, rise time and flat time, are set as 8 $\mu$s and 1$\mu$s, respectively. 
As shown in Fig.~\ref{fig:Filter}, 
the trapezoidal filter converts the raw charge pulse to a trapezoid pulse in which 
the height of the trapezoid indicates the amplitude of the raw pulse.

\begin{figure}[!htb]
\includegraphics
[width=1.0\hsize]
{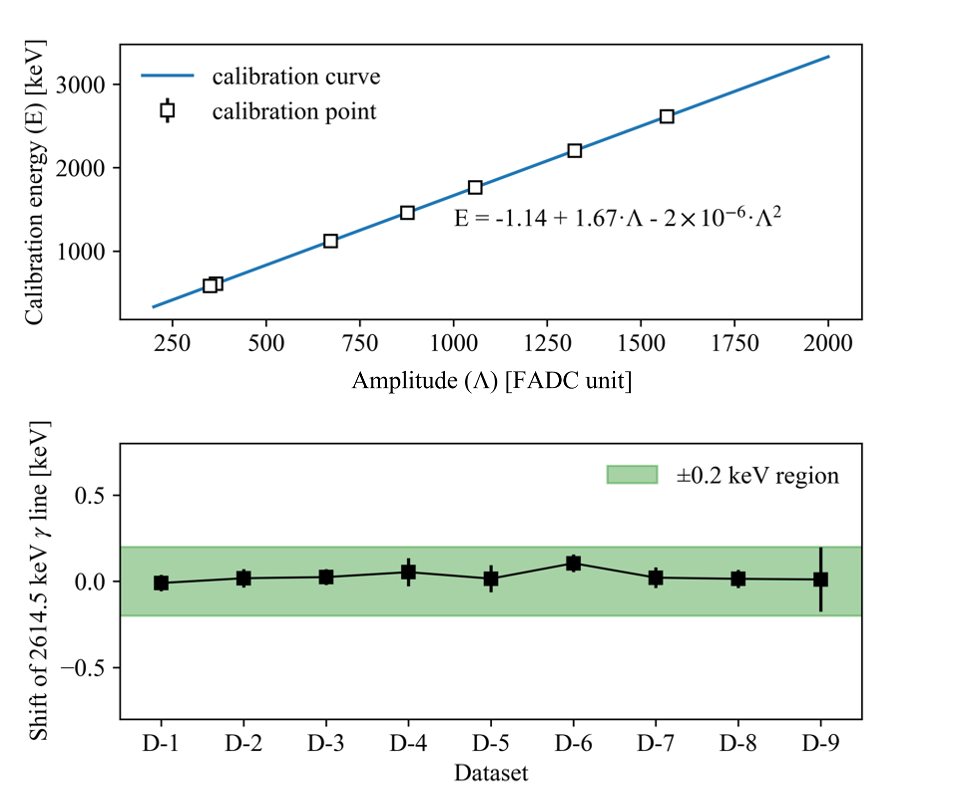}
\caption{\label{fig:ECal} 
% The energy calibration of one of 9 datasets.
Top panel: The energy calibration of one of 9 datasets,
the amplitude ($\Lambda$) is converted to the calibration energy (E) 
via a second order polynomial: 
$E=k_0\cdot\Lambda^2+k_1\cdot\Lambda+k_2$;
Bottom panel: shift of 2614.5 keV
peak of $^{208}$Tl during data taking.}
\end{figure}

\begin{figure}[!htb]
\includegraphics
[width=1.0\hsize]
{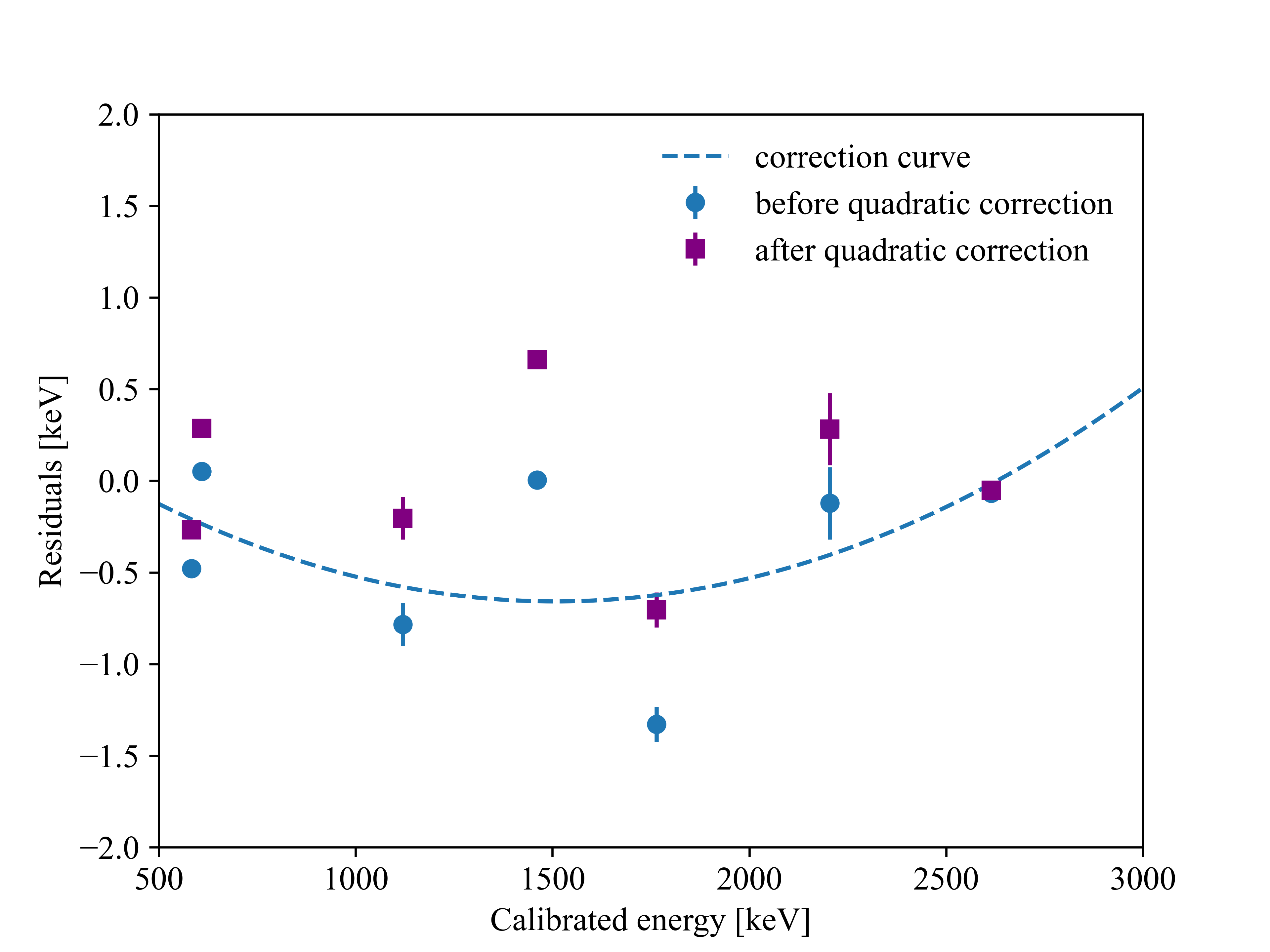}
\caption{\label{fig:EQC} 
% Quadratic correction of the combined data, 
% the residuals of the calibrated energy are fitted with a quadratic function.
Residuals of the
reconstructed peak energy from calibration fit and 
expected peak energy, before correction (blue circles) 
and after introducing the nonlinearity correction in the 
calibration (purple squares)}
\end{figure}

Energy calibrations are performed in each dataset using characteristic gamma peaks
from primordial radionuclides in the detector and its surrounding materials. 
Seven peaks from $^{208}$Tl (583.3 keV, 2614.5 keV), 
$^{214}$Bi (609.3 keV, 1120.3 keV, 1764.5 keV, 2204.1 keV), 
and $^{40}$K (1460.8 keV) are used in calibrations.
Each peak is fitted with a Gaussian function coupled with a linear background 
to determine the peak position. 
A second-order polynomial is used to convert amplitude to energy. 
Top panel of Fig.~\ref{fig:ECal} shows the calibration curve of one dataset.
The stability of the detector is evaluated via 
the shift of 2614.5 keV line 
as shown in the 
bottom panel of Fig.~\ref{fig:ECal}. 
The shift is within 0.2 keV
during the data taking, 
indicating that the detector is under a stable operation. 
After combining all calibrated datasets, 
a maximum 1.5 keV residual is found in the in 
the characteristic gamma peaks.
And a nonlinearity correction [\onlinecite{bib:29}] 
is adopted to reduce the residuals of fitted energy.
This correction is applied in the combined data 
to reduce the statistical uncertainties in each gamma peak, 
the corrected energies are shown in Fig.~\ref{fig:EQC}. 
After the correction, the maximum residual
(0.7 keV in $^{40}$K 1460.8 keV line) 
is adopted as a systematic uncertainty 
in the energy reconstruction of 
$0\nu\beta\beta$ events.
% is below 0.7 keV 
% for all seven characteristic gamma peaks.

\subsection{\label{sec:3.2}Pulse shape discrimination}

Since the ranges in a germanium crystal of the two electrons of a 
0$\nu\beta\beta$ decay event are of the order of 1 mm, 
0$\nu\beta\beta$ events are typical single-site events (SSEs). 
High energy gamma rays are expected to deposit their energies at 
multiple sites featuring the so-called multi-site events (MSEs).

\begin{figure}[!htb]
\includegraphics
[width=0.9\hsize]
{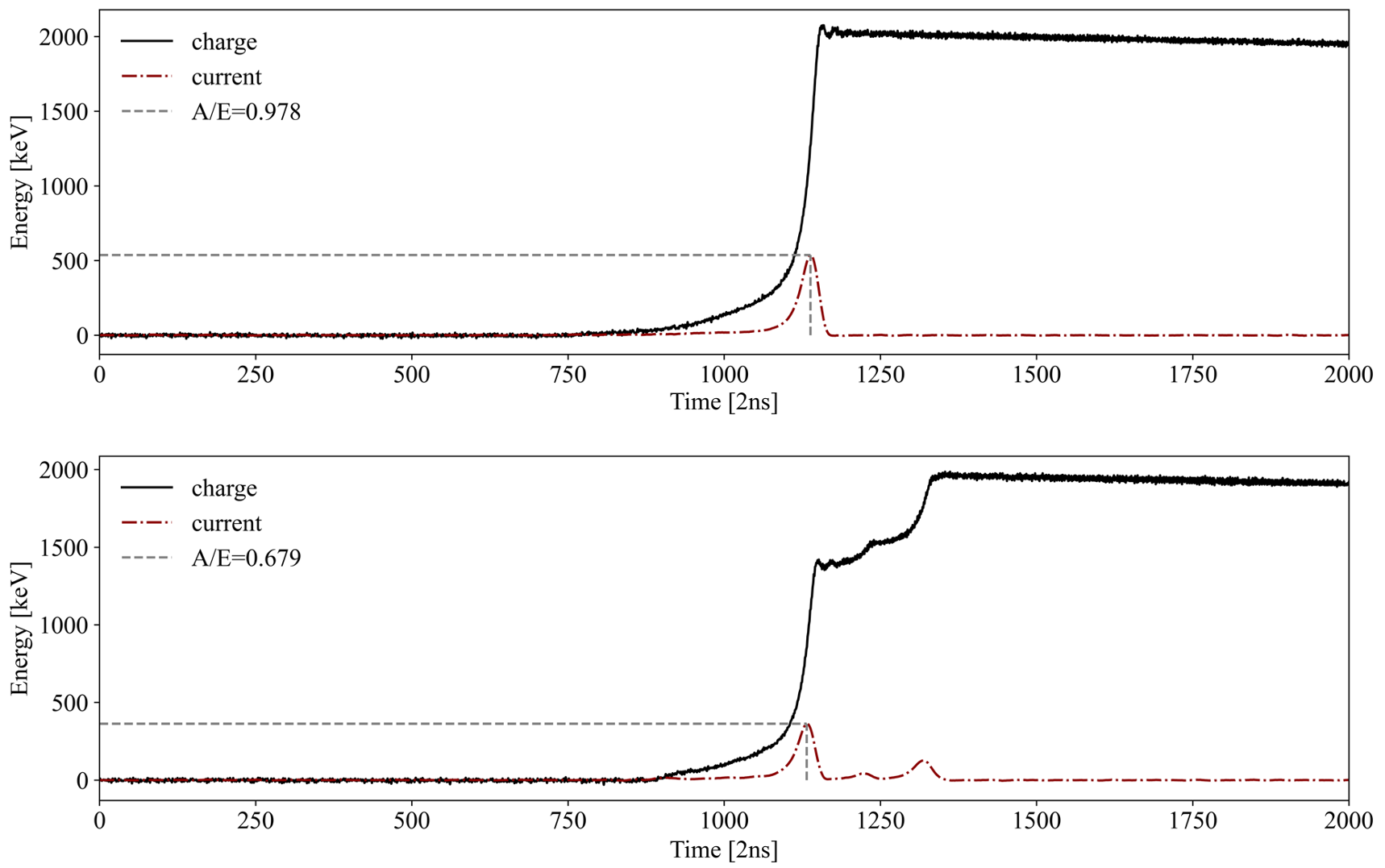}
\caption{\label{fig:SMSE} Typical charge and current pulses of a SSE / MSE, 
the current pulses have been rescaled for demonstration.}
\end{figure}

A pulse shape discrimination (PSD) method can be used to 
discriminate between single-site events and multi-site events 
in a BEGe detector [\onlinecite{bib:30},\onlinecite{bib:31}].
The PSD method relies on the A/E parameter, 
in which A is the maximum amplitude of the current pulse and 
E is the reconstructed energy. 
The current pulse is extracted from the charge pulse by a moving average differential filter. 
Fig.~\ref{fig:SMSE} shows charge and current pulses of a typical SSE and MSE, respectively. 
SSEs deposit energies in a small range of area. The current of a SSE has one peak, 
with an amplitude A proportional to the energy E. 
MSEs deposit energies in multiple detector positions, 
leading to multi-peaks in current pulses and lower A/E values than those of SSEs.

A $^{228}$Th calibration experiment is conducted to determine the acceptance region of the A/E cut, 
the detector is irradiated by a $^{228}$Th source to create double escape events from 
$^{208}$Tl 2614.5 keV $\gamma$-rays. Events in the 1592.5 keV double escape peak (DEP) 
have a similar profile as the 0$\nu\beta\beta$ events [\onlinecite{bib:30}] 
and therefore are used as proxies of SSEs.
Events in the single escape peak (SEP) are typical two-site events and are used as proxies of MSEs. 
The A/E distribution of DEP events is fitted with a Gaussian function to determine 
the mean ($\mu_{A⁄E}^{SSE}$) and standard deviation
($\sigma_{A⁄E}^{SSE}$) of A/E parameters for SSEs. 
The acceptance region of the A/E cut is set to 
($\mu_{A⁄E}^{SSE}$$\pm$5$\sigma_{A⁄E}^{SSE}$) and leads to a 
93\% survival rate of the DEP events, 
and a 5\% survival rate of the SEP events.

\begin{figure}[htb]
\includegraphics
[width=0.9\hsize]
{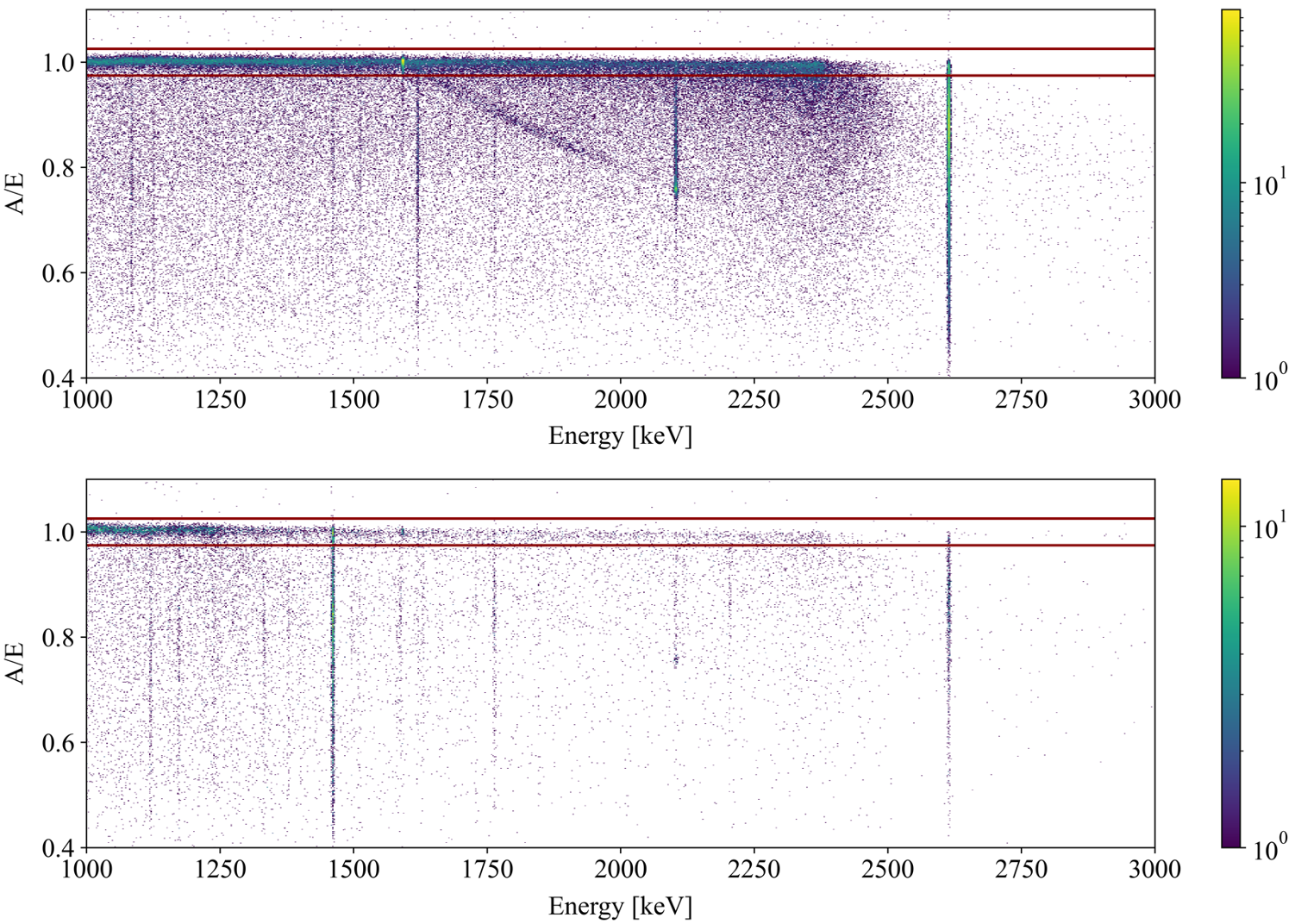}
\caption{\label{fig:AEMap} A/E versus energy distributions of the
$^{228}$Th calibration data (top) and the 186.4 kg$\cdot$day exposure data (down). 
The red lines indicate the A/E acceptance region for SSEs.}
\end{figure}

Fig.~\ref{fig:AEMap} shows A/E discriminations applied in 
the $^{228}$Th calibration data (top) and 
the 186.4 kg$\cdot$day exposure data (down). 
The low A/E cut removes MSEs. 
Events rejected by the high A/E cut are likely to be $\alpha$ events 
originating from the surface contamination and the p+ electrode. 

% \begin{figure}[htb]
% \includegraphics
% [width=0.9\hsize]
% {FigAEStability.png}
% \caption{\label{fig:AEStability} A/E cut stability during 
% the data taking, the survival fractions are fitted with a
% flat line, the $\chi^2$ and 
% the number of degree of freedom ($NDF$) are used to 
% demonstrate the goodness of the fit.}
% \end{figure}

% The stability of the A/E cut is 
% monitored using the survival fraction (SF) of the cut. 
% The survival fractions of events in $>$1000 keV energy region 
% and 1800$\sim$2200 keV energy region are plotted in 
% Fig.~\ref{fig:AEStability}.
% A flat line is used to fit the data to check the 
% stability of the cut via a least square method. 
% The $\chi^2/NDF$ of those two types of events are 
% 18.10/26 and 19.05/26, 
% correspond to p-values of 0.87 and 0.83, respectively. 
% Despite the statistical uncertainties in the 
% 1800$\sim$2200 keV events, 
% the SFs of those two types of events are well fitted by 
% a flat line, indicates the performance of A/E cut is stable 
% during the data taken.

Survival fraction (SF) of the A/E cut in 1800$\sim$2200 keV
energy region is used to evaluate the stability of the cut.
The SF is fitted with a flat line via the least square method. 
The $\chi^2/$\textit{(degree of freedom)} of 
the fit is 19.05/26, indicating that the performance 
of A/E cut is stable during data taking.

\begin{figure}[htb]
\includegraphics
[width=1.0\hsize]
{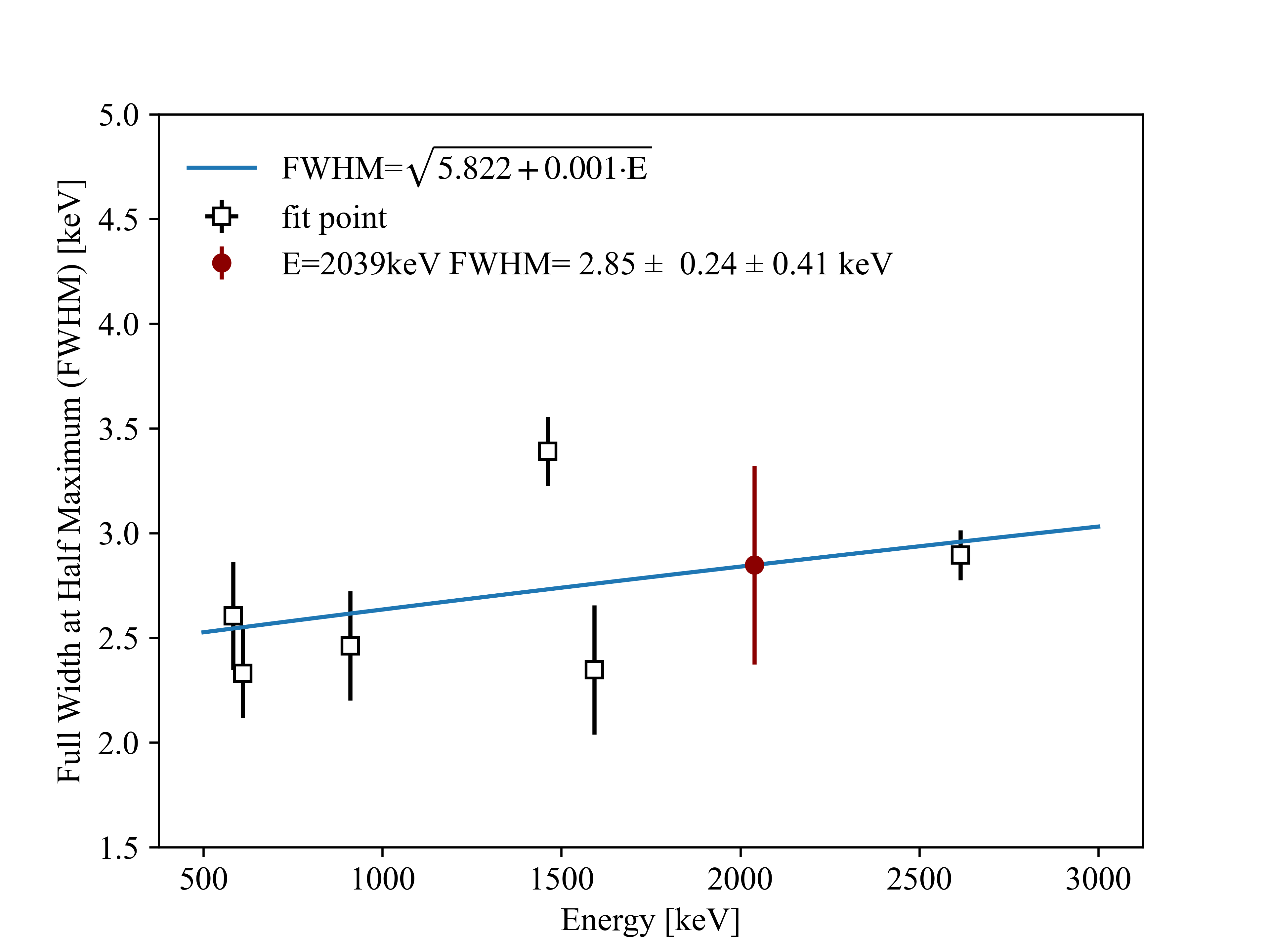}
\caption{\label{fig:FWHM} Energy resolutions of single-site events.}
\end{figure}

Selected SSEs in the 186.4 kg$\cdot$day exposure data 
are used to evaluate the 
energy resolution of 0$\nu\beta\beta$ signals, 
indicated as the full width at half maximum (FWHM),
as shown in Fig.~\ref{fig:FWHM}. 
Gamma peaks from $^{208}$Tl (583.3 keV, 1592.5 keV, 2614.5 keV), 
$^{214}$Bi (609.3 keV), $^{228}$Ac (911.2 keV) and $^{40}$K (1460.8 keV) 
are used to calculate the FWHM at 2039 keV.
The FWHMs of the six peaks are fitted with a function 
FWHM = $\sqrt{a+bE}$, 
and the interpolation of FWHM at 2039 keV is 2.85 keV.
Uncertainties of the result mainly originate from two aspects:
\begin{itemize}
 \item[(1)]
 Uncertainties from FWHMs of the selected characteristic gamma peaks 
and their effects on curve parameters ($a$, $b$):
The uncertainties are calculated within the standard chi-square fitting and 
error propagation techniques. The combined uncertainty is $\pm$0.24 keV. 
 \item[(2)] 
 Choice of characteristic gamma peaks: 
 Systematic effects are taken as deviations of results due to 
 the choice of gamma peaks. 
 The FWHM curve is refitted without one of the six aforementioned peaks, 
 and the maximum deviations in the FWHM at 2039 keV is 0.41 keV 
 when the 2614 keV peak is excluded.
\end{itemize}

Combining both uncertainties, 
the FWHM at 2039 keV is given as (2.85$\pm$0.48) keV

\subsection{\label{sec:3.3} Efficiency calibration}

The total 0$\nu\beta\beta$ signal efficiency consists of: 
(i) the efficiency of the data quality cut ($\varepsilon_{QC}$), 
(ii) the efficiency of the two electrons emitted from 0$\nu\beta\beta$ decay 
deposits all energy in the active volume of the detector ($\varepsilon_{fed}$), and 
(iii) the efficiency of PSD ($\varepsilon_{PSD}$). 
The trigger rate during data taken is measured to be 0.005 Hz, 
the dead time is negligible and not considered in the efficiency.

The efficiency loss due to the noise cut is negligible as 
a physical event can be rejected by the noise cut only when 
it is overlapped with a burst of noise event. And
the coincidences of those two events are negligible
because of the low trigger rate.
The efficiency of the data quality cut is calculated by recorded physical events, 
given as (94.37$\pm$0.49)\% 
where the error is the statistical uncertainty of the recorded events.

The n+ electrode on the side and top surface of the detector forms an inactive region, 
known as the dead layer, reducing the active volume of the detector. 
The dead layer of (1.18$\pm$0.10) mm and (0.17$\pm$0.10) mm for side and top surfaces 
have been measured in our previous work [\onlinecite{bib:32}-\onlinecite{bib:34}] 
and gives a (91.1$\pm$0.96)\% active volume of the crystal.

\begin{figure*}[htbp]
\includegraphics
[width=0.8\hsize]
{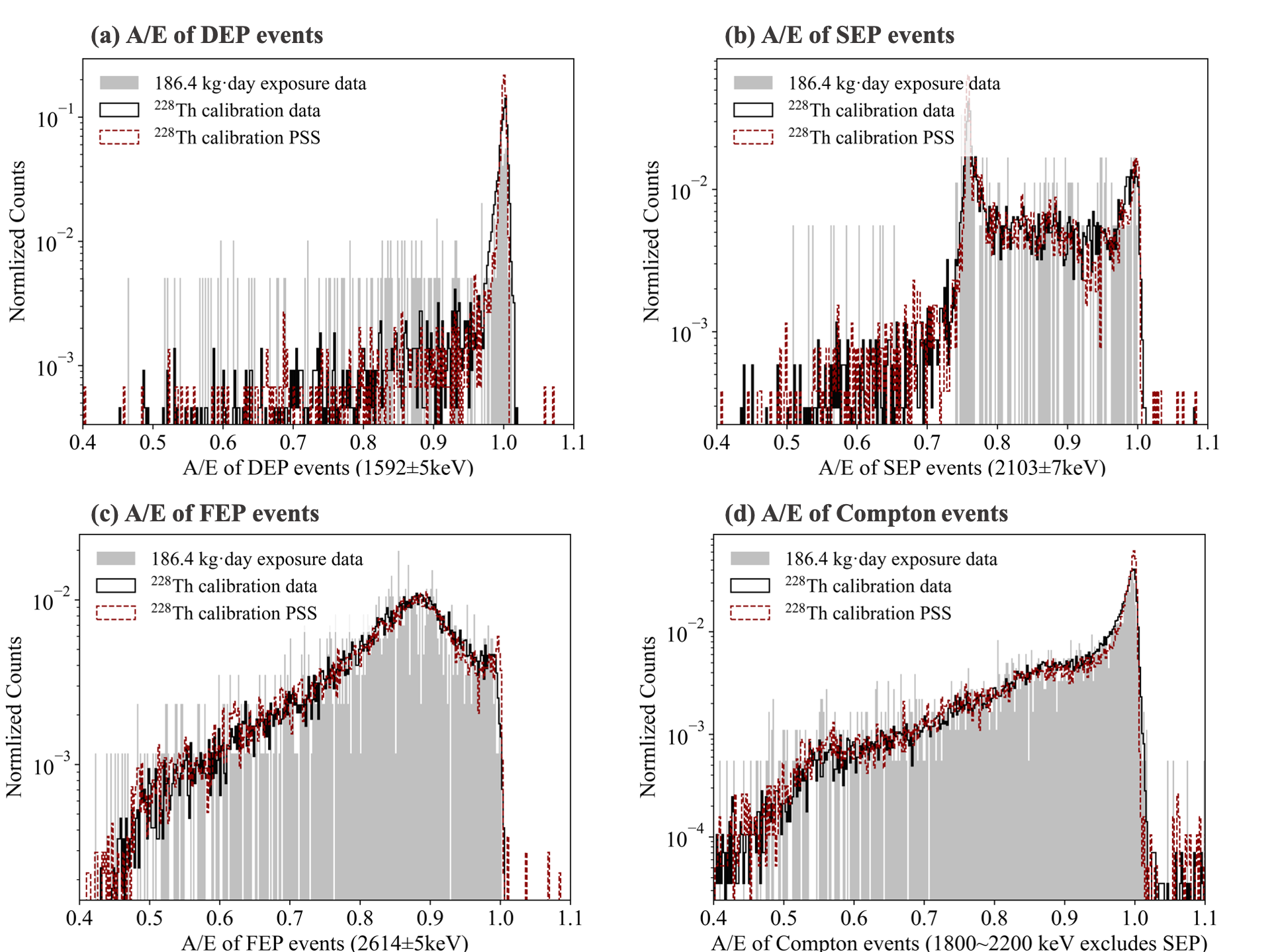}
\caption{\label{fig:PSS} A/E distributions derived from PSS, 
$^{228}$Th calibration and 186.4 kg$\cdot$day exposure data, 
for (a) the DEP events (1592$\pm$5 keV), 
(b) the SEP events (2103$\pm$7 keV), 
(c) the FEP events (2614$\pm$5 keV), and 
(d) the Compton events (1800$\sim$2200 keV excludes SEP). 
All counts are normalized for comparison.}
\end{figure*}

The probability of 0$\nu\beta\beta$ events deposit all energy in the 
active volume of the detector ($\varepsilon_{fed}$) is calculated by Monte Carlo simulations 
via a Geant4 based simulation toolkit SAGE [\onlinecite{bib:35}]. 
The 0$\nu\beta\beta$ decays are uniformly sampled in the germanium crystal, 
events with full energy deposited in the active region are counted to calculate the efficiency.
The efficiency ($\varepsilon_{fed}$) is (86.71$\pm$0.84)\% 
where the error is derived from variations of the dead-layer thickness: 
efficiencies are calculated for the top dead-layer thickness ranging from 0.07$\sim$0.27mm 
and the side dead-layer thickness ranging from 1.08$\sim$1.28mm, 
the maximum deviation on results is counted as the uncertainty. 
The $\varepsilon_{fed}$ is lower than the active volume because 
the two electrons in 0$\nu\beta\beta$ decay may lose their energy in germanium by bremsstrahlung.

The $^{228}$Th calibration data and pulse shape simulations (PSS) are used to determine the PSD efficiency. 
The PSS of the $^{228}$Th calibration data is conducted within 
a pulse shape simulation module of the SAGE toolkit [\onlinecite{bib:36}]. 
The A/E distributions derived from the PSS, the $^{228}$Th calibration data 
and the 186.4 kg$\cdot$day exposure data are compared in Fig.~\ref{fig:PSS}.
Events from the DEP, the SEP, the full energy peak (FEP) and the Compton flat (1800$\sim$2200 keV) of 
$^{208}$Tl 2614.5 keV $\gamma$ lines are selected for comparison. 
Table.~\ref{tab:PSS} lists the A/E cut removal and survival fractions of the simulation and the calibration data.

% \begin{figure}[htbp]
% \includegraphics
% [width=1.0\hsize]
% {FigPSS.png}
% \caption{\label{fig:PSS} A/E distributions derived from PSS, 
% $^{228}$Th calibration and 186.4 kg$\cdot$day exposure data, 
% for (a) the DEP events (1592$\pm$5 keV), 
% (b) the SEP events (2103$\pm$7 keV), 
% (c) the FEP events (2614$\pm$5 keV), and 
% (d) the Compton events (1800$\sim$2200 keV excludes SEP). 
% All counts are normalized for comparison.}
% \end{figure}

\begin{table*}[htbp]
\caption{\label{tab:PSS}Removal fractions by the low A/E cut 
and high A/E cut and total survival fractions applying both cuts 
in $^{228}$Th calibration data and pulse shape simulation data.}
\begin{ruledtabular}
 \renewcommand\arraystretch{1.5}
\begin{tabular}{cccc}
 Region&Low A/E cut&High A/E cut&Survival fraction\\
 &A/E $<$ 0.975&A/E $>$ 1.025&0.975 $<$ A/E $<$ 1.025 \\ \hline
 $^{228}$Th calibration data \\
 DEP 1592.5 keV & 6.76$\pm$0.67\% & 0.12$\pm$0.09\% & 93.12$\pm$3.32\% \\
 $^{228}$Th calibration PSS \\
 DEP 1592.5 keV & 6.55\% & 1.43\% & 92.03\% \\
 0$\nu\beta\beta$ events PSS \\
 Q$_{\beta\beta}$ 2039 keV & 6.99\% & 3.77\% & 89.23\% \\
\end{tabular}
\end{ruledtabular}
\end{table*}

\begin{table*}[htbp]
 \caption{\label{tab:eff}Uncertainties of 0$\nu\beta\beta$ signal efficiency 
 and their compositions, the combined efficiency and 
 its uncertainty are listed in the last column.}
 \begin{ruledtabular}
   \renewcommand\arraystretch{1.5}
 % \begin{tabular}{p{3.4cm}p{9cm}c}
 \begin{tabular}{ccc}
   Sources of Efficiency&Sources of Uncertainties&Value / [type] \\ \hline
   Quality Cut & \multirow{2}{*}{Statistical uncertainty of recorded events} 
   & \multirow{2}{*}{$\pm$0.49\% [stat]}\\
   $\varepsilon_{QC}$=94.37\% && \\ \hline
   0$\nu\beta\beta$ events full energy deposition &\multirow{2}{*}{Uncertainty on dead-layer thickness}
   & \multirow{2}{*}{$\pm$0.84\% [sys]} \\
   $\varepsilon_{fed}$=86.71\% & & \\ \hline
   Pulse shape discrimination & Low A/E cut removal fraction of $^{228}$Th calibration data & $\pm$0.67\% [stat]\\
   $\varepsilon_{PSD}$=89.47\%& Differences between 0$\nu\beta\beta$ and DEP events & $\pm$0.44\% [sys] \\
   & Differences between calibration and physics data & $\pm$0.91\% [sys] \\
   & Variations on PSS parameters & $\pm$0.97\% [sys] \\
   & Maximum discrepancy between experiment and PSS & $\pm$1.31\% [sys] \\ \hline
   \multirow{2}{*}{Combined efficiency} & Efficiency = $\varepsilon_{QC}\cdot\varepsilon_{fed}\cdot\varepsilon_{PSD}$ = 73.21\% \\
   &Uncertainty = $\pm$1.84\% & \\
 \end{tabular}
 \end{ruledtabular}
\end{table*}

0$\nu\beta\beta$ events and DEP events are both typical SSEs but have 
different locations in the detector, 0$\nu\beta\beta$ events are 
homogeneously distributed while DEP events are dominantly 
located at the corners. Therefore, the PSD efficiency of 0$\nu\beta\beta$ events 
($\varepsilon_{PSD}$) is calculated by a similar way of G\textsc{erda} [\onlinecite{bib:37}]: 
the removal fraction of the low A/E cut is adopted from the 
$^{228}$Th calibration data as the low A/E cut only removes MSEs and 
0$\nu\beta\beta$ events and DEP events are both typical SSEs. 
The removal fraction of the high A/E cut is adopted from the 
pulse shape simulation of 0$\nu\beta\beta$ events. 
The calculation gives a PSD efficiency of 89.47\%, 
similar to the result derived from the PSS (89.23\%)

Statistical and systematic uncertainties of the PSD efficiency 
mainly consist of four parts:
\begin{itemize}
 \item[(1)]
 The statistical uncertainty of the low A/E cut fraction of DEP events, $\pm$0.67\%; 
 \item[(2)]
 The systematic uncertainty due to differences 
 between 0$\nu\beta\beta$ and DEP events: 
 the discrepancy between the 
 removal fraction of the simulated DEP events and 
 0$\nu\beta\beta$ events by the low A/E cut
%  low A/E cut removal fraction of the DEP events and 
%  the simulated 0$\nu\beta\beta$ events 
 is counted as the uncertainty, $\pm$0.44\%;
 \item[(3)]
 The systematic uncertainty due to the residual differences 
 between calibration and physics data:
 the survival fraction of $^{208}$Tl 
 2614.5 keV peak is used to compute the uncertainty. 
 The discrepancy between 
 the calibration data (5.39\%$\pm$0.2\%)
 and physical data (5.48\%$\pm$0.8\%)
 is (0.09\%$\pm$0.82\%).
%  To stand on the safe side, 
 The upper limit of the discrepancy is adopted 
 as the systematic uncertainty, $\pm$0.91\%;
 \item[(4)] 
 The systematic uncertainty of PSS: identical analyses are performed on varies PSS parameters, 
 and the maximum deviation on results is adopted as one systematic uncertainty ($\pm$0.97\%). 
 The maximum deviation between the $^{228}$Th calibration data and the PSS in Table.~\ref{tab:PSS} (±1.31\%) 
 is added as the other systematic uncertainty. The combined systematic uncertainty is $\pm$1.63\%.
\end{itemize}

Combining the statistical and systematic uncertainties, the $\varepsilon_{PSD}$ 
is given as (89.47$\pm$2.03)\%.

Compositions of the 0$\nu\beta\beta$ signal efficiency and 
their uncertainties are listed in Table.~\ref{tab:eff}. The total efficiency is the product of the 
$\varepsilon_{QC}$, $\varepsilon_{fed}$, and $\varepsilon_{PSD}$,
i.e. (73.21$\pm$1.84)\%.

\subsection{\label{sec:3.4} Background model}

Background spectra of different radioactive isotopes in different components of 
the detector setup are simulated using the SAGE toolkit. 
Table.~\ref{tab:bkg} lists all the simulated background sources and components. 
In our simulation, secular equilibrium is assumed in $^{238}$U and $^{232}$Th decay chain and 
all background sources are assumed to be uniformly distributed in their components.
Due to the low muon flux in CJPL-I [\onlinecite{bib:25}], 
backgrounds from muons and their secondary particles are negligible 
(less than 1$\times$10$^{-6}$ counts/keV/kg/day (cpkkd)). 
Neutrons are also negligible after shields of 
a 1 m polyethylene wall and a 20 cm borated polyethylene absorber. 
Therefore, they are not considered in the simulation. 
The 2$\nu\beta\beta$ decays of $^{76}$Ge are considered assuming 
a half-life of 2.1$\times$10$^{21}$ yr [\onlinecite{bib:38}].

\begin{table}
 \caption{\label{tab:bkg}Simulated background components and their 
 contributions (R$_{0\nu\beta\beta}$) in the 0$\nu\beta\beta$ signal region.}
 \begin{ruledtabular}
   \renewcommand\arraystretch{1.5}
 \begin{tabular}{ccc}
   \multirow{2}{*}{Sources}&\multirow{2}{*}{Components}&\multirow{2}{*}{R$_{0\nu\beta\beta}$}\\
   &&\\ \hline
   Cosmogenic & $^{68}$Ge, $^{60}$Co, $^{54}$Mn, $^{65}$Zn in Ge & 8.6\% \\
   isotopes& $^{60}$Co in Copper & 2.1\% \\ \hline
   \multirow{5}{*}{$^{238}$U chain}&Crystal holder&\multirow{5}{*}{15.1\%}\\
   &Signal pin, Electronics&\\
   &Vacuum Cup&\\
   &Outer Shield&\\ 
   &$^{222}$Rn&\\ \hline
   \multirow{4}{*}{$^{232}$Th chain}&Crystal holder&\multirow{4}{*}{74.2\%}\\
   &Electronics&\\
   &Vacuum Cup& \\
   &Outer Shield&\\ \hline
   \multirow{4}{*}{$^{40}$K}&Crystal holder&\multirow{4}{*}{0\%}\\
   &Electronics&\\
   &Vacuum Cup& \\
   &Outer Shield&\\
 \end{tabular}
 \end{ruledtabular}
\end{table}

\begin{figure}[htbp]
\includegraphics
[width=1.0\hsize]
{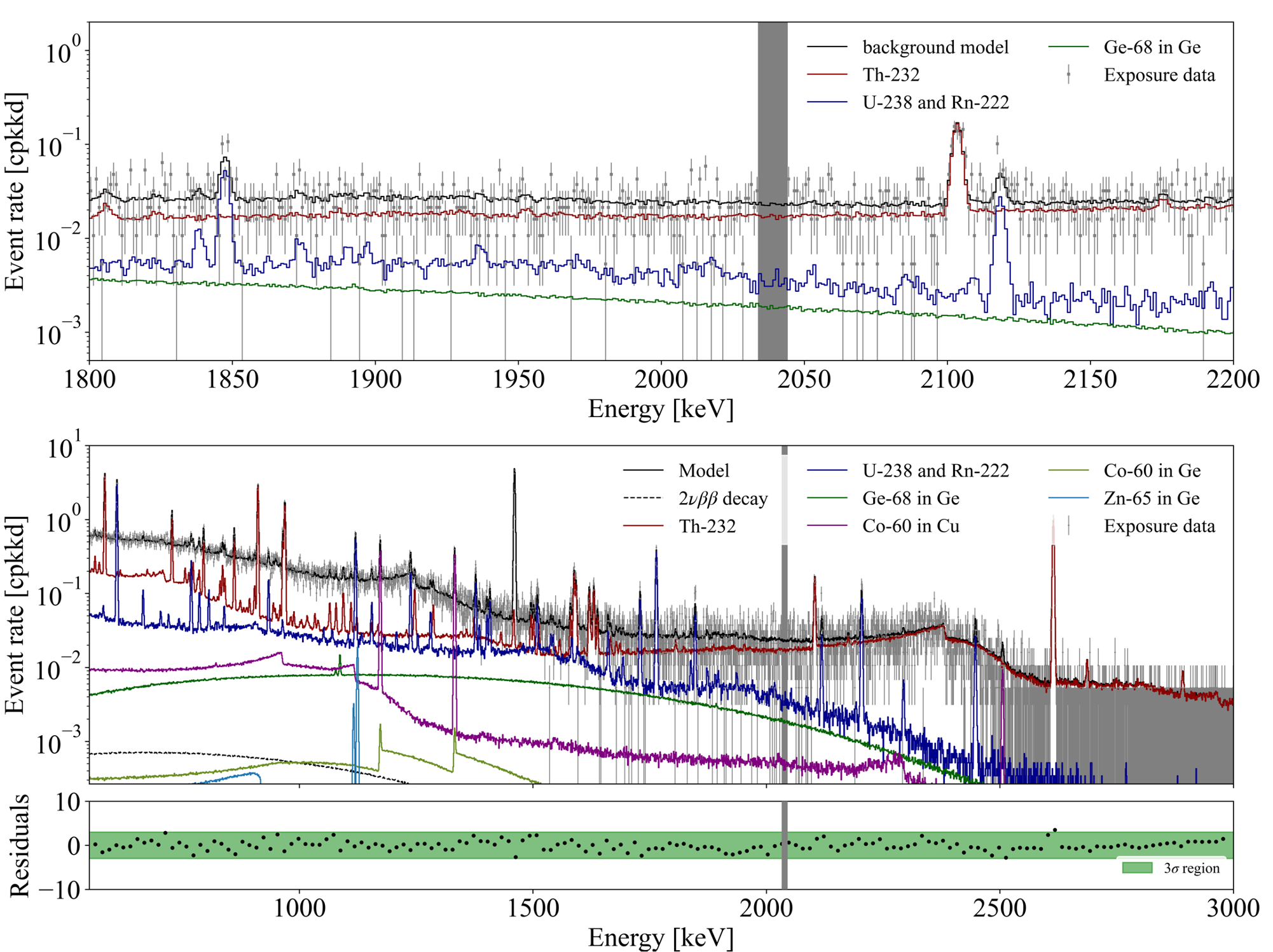}
\caption{\label{fig:bkgmodel} Background model and its decompositions. 
Top panel shows spectra in 1800$\sim$2200 keV. 
Bottom panel shows spectra in 550$\sim$3000 keV. 
The simulated spectra are fitted with exposure data prior to the PSD, 
and the normalized residuals are shown under the spectra, 
the 3-$\sigma$ band is marked in green. 
The blind regions (Q$_{\beta\beta}\pm$5 keV) are labeled in gray. 
The black dotted line is the expected 2$\nu\beta\beta$ spectrum 
assuming a half-life of 2.1$\times$10$^{21}$ yr [\onlinecite{bib:38}].}
\end{figure}

A background model (Fig.~\ref{fig:bkgmodel}) is obtained by 
fitting the 186.4 kg$\cdot$day spectrum with simulated spectra 
in 550$\sim$3000 keV energy range, using the maximum likelihood method. 
The simulated spectra are convolved with an energy resolution function 
derived from fitting the FWHMs of the prominent gamma peaks in the
spectrum prior to the PSD. Contributions of background sources in the 
0$\nu\beta\beta$ signal region are determined from the background model 
and are listed in Table.~\ref{tab:bkg}.

In the 0$\nu\beta\beta$ signal region (Q$_{\beta\beta}\pm$ 5keV), 
89\% of the background are from radionuclides in the $^{232}$Th and $^{238}$U decay chains 
according to the background model. 
A background of 2.29$\times$10$^{-2}$ cpkkd in Q$_{\beta\beta}\pm$5keV region 
projected by the background model agrees well with (2.13$\pm$0.3)$\times$10$^{-2}$ cpkkd 
calculated from the exposure data after unblinding.

\section{\label{sec:4}Results and Discussion}
\begin{figure}[htb]
\includegraphics
[width=1.0\hsize]
{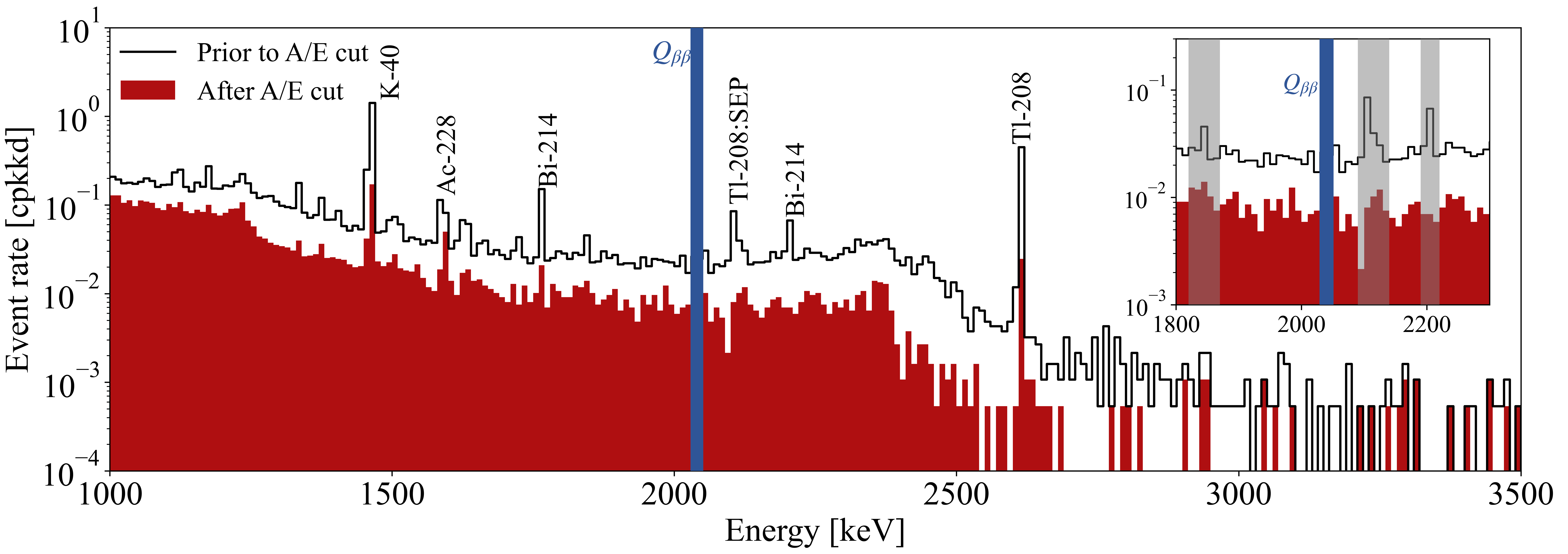}
\caption{\label{fig:bkgspec} Spectra of the 186.4 kg$\cdot$day data, 
the main gamma lines are labeled in the spectrum prior to the PSD. 
The inset shows the same spectra in the background estimation window, 
which spans 1800$\sim$2300 keV, with regions excluded due to $\gamma$ backgrounds 
shaded in gray and the 10 keV window centered at Q$_{\beta\beta}$ shaded in blue.}
\end{figure}

Fig.~\ref{fig:bkgspec} shows the measured energy spectra above 1000 keV 
for the 186.4 kg$\cdot$day exposure data. 
% The spectrum shown in black has only quality cuts applied. 
% The spectrum in red also has PSD cut (A/E cut) applied. 
Spectra shown in black and red are prior to and after the PSD, respectively.
Spectra in the energy region of 1800$\sim$2300 keV are used to estimate the background in the 
0$\nu\beta\beta$ region of interest (ROI). Gamma peaks identified by the background model, 
as indicated by gray shading in the inset of Fig.~\ref{fig:bkgspec}, 
are excluded. Additionally, a $\pm$5 keV wide window centered at 
Q$_{\beta\beta}$ is excluded, as indicated by the blue shaded region. 
Prior to and after PSD, the estimated background in the ROI from the 
resulting 420 keV window is (2.45$\pm$0.06)$\times$10$^{-2}$ cpkkd 
and (0.64$\pm$0.03)$\times$10$^{-2}$ cpkkd, respectively. 
The background in the ROI is reduced by a factor of 3.79 
after applying the PSD method.

The exposure data after all cuts are used to analyze the 
0$\nu\beta\beta$ decay of $^{76}$Ge. The half-life of 
$^{76}$Ge neutrinoless double-beta decay can be calculated by 
Eq.~\ref{eq:1}:

\begin{equation}
T_{1/2}^{0\nu}=\frac{\textup{ln}2\cdot N_A\cdot f_{76}\cdot m\cdot T\cdot \varepsilon_{total}}{N^{0\nu}\cdot M}
\label{eq:1}
\end{equation}

Where $m\cdot T$ is the exposure, 
$\varepsilon_{total}$ is the total efficiency defined in Sec~\ref{sec:3.3}, 
${N^{0\nu}}$ is the number of observed 0$\nu\beta\beta$ signal events, 
$M$ the molar mass of natural Ge, 
$N_A$ is the Avogadro’s constant, 
$f$ the fraction of $^{76}$Ge atoms in the natural germanium detector.

\begin{figure}[htbp]
\includegraphics
[width=1.0\hsize]
{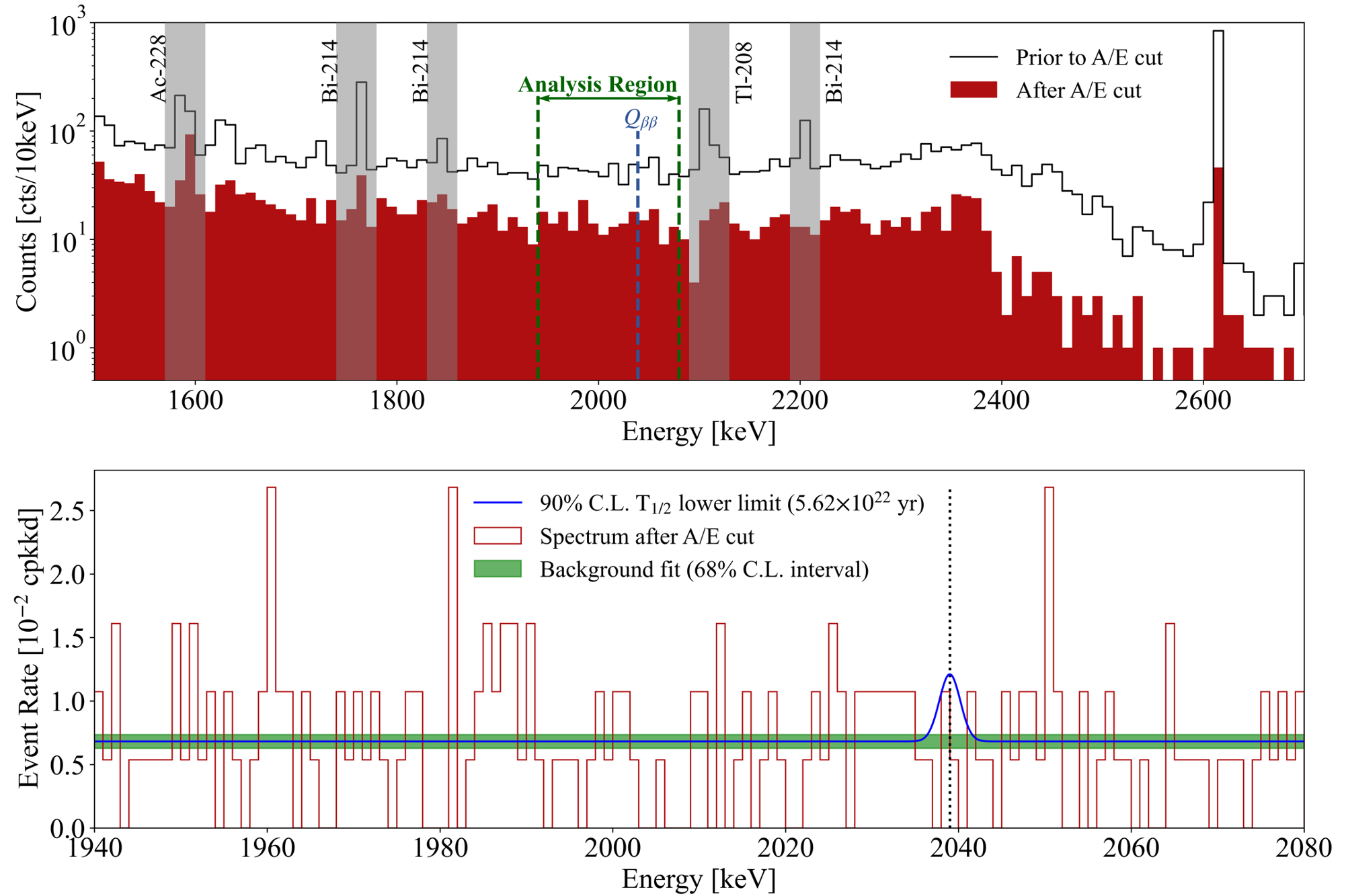}
\caption{\label{fig:ROIspec} Top panel: energy spectra of events 
before and after the A/E cut, 
green lines indicate the analysis region; 
Bottom panel: the spectrum of events survive all cuts in the analysis region, 
the blue line is the best fit background 
added with signal spectrum corresponding to 
the 90\%C.L. lower limit of the 0$\nu\beta\beta$ half-life, 
the green area is the 68\%C.L. interval of the background fit result.}
\end{figure}

Spectra of the 1940$\sim$2080keV analysis region are shown in Fig.~\ref{fig:ROIspec}. 
After unblinding, 178 events survive all data cuts, and eight events are found 
in the ROI (Q$_{\beta\beta} \pm$3$\sigma_{\beta\beta}$).

The 0$\nu\beta\beta$ decay signal is analyzed using 
an unbinned extended profile likelihood method [\onlinecite{bib:39}]. 
As predicted by the background model, no background peak is identified, 
and a flat background is assumed in the analysis region. 
For the signal, a Gaussian distribution centered 
at the Q$_{\beta\beta}$ with a width corresponding 
to the energy resolution is considered.
The likelihood function is then given by:

\begin{eqnarray}
f(E|b,N^{0\nu })=&&\frac{1}{\triangle E\cdot b+N^{0\nu } } \\ \nonumber
&&\times\left ( b+\frac{N^{0\nu }}{\sqrt{2\pi}\cdot \sigma}e^{-\frac{(E-Q_{\beta\beta})^2}{2\sigma^{2}}} \right )
\label{eq:2}
\end{eqnarray}
\begin{eqnarray}
L(b,N^{0\nu})=&&\frac{(\triangle E\cdot b+N^{0\nu})^N\cdot e^{-(\triangle E\cdot b+N^{0\nu})}}{N!} \\ \nonumber
&&\times\prod_{i=1}^{N}f(E_i|b,N^{0\nu})
\label{eq:3}
\end{eqnarray}

Where N is the total events number in the analysis region, 
b is the background rate (cts/keV) and 
$\triangle$E is the width of the analysis region, 
$N^{0\nu}$ the observed 0$\nu\beta\beta$ events, 
E the energy of recorded events, 
$\sigma$ is the energy resolution at $Q_{\beta\beta}$. 
$f(E|b,N^{0\nu})$ is the probability density function (pdf) of one single event. 
The likelihood function $L(b,N^{0\nu})$ is the product of the pdf of each event and extended with the Poisson term.

When fitting the likelihood function L, 
parameter b and $N^{0\nu}$ are bound to positive values. 
And a test statistic based on profile likelihood is used to calculate the confidence interval. 
The probability distributions of the test statistic are computed using the Monte Carlo method.

The unbinned profile likelihood analysis yields a best-fit 
background of 1.27 cts/keV and no indication for signal. 
The lower limit for 0$\nu\beta\beta$ decay half-life is set to:

\begin{eqnarray}
T_{1/2}^{0\nu}\geq 5.62\times 10^{22} \textup{ yr at 90\%C.L.} 
\label{eq:4}
\end{eqnarray}

The corresponding 90\%C.L. upper limit of 
the 0$\nu\beta\beta$ signal strength is 2.99 events. 
Uncertainties of the energy calibration (2039$\pm$0.7 keV) and 
the energy resolution (2.85$\pm$0.48 keV) are considered by 
folding them into the profile likelihood function through 
additional nuisance parameters constrained by Gaussian probability distributions. 
Uncertainties of the efficiency and the exposure are 
considered by propagating them through Eq.~\ref{eq:1}. 
The overall effect of all uncertainties on the half-life limit is about 2.67\%.

The upper limit on the effective Majorana neutrino mass $m_{\beta\beta}$
is derived by:

\begin{equation}
(T_{1/2}^{0\nu})^{-1}=G^{0\nu}\left | g_A^2\cdot M_{0\nu} 
\right |^2\left (\frac{m_{\beta\beta}}{m_e}  \right )^2
\label{eq:5}
\end{equation}
\begin{equation}
m_{\beta\beta}\leq [4.6,10.3] \textup{ eV}
\label{eq:6}
\end{equation}

The upper and lower values are obtained by 
using nuclear matrix elements 
$M_{0\nu}$ from [\onlinecite{bib:40}] and [\onlinecite{bib:41}], respectively, 
the coupling constant $g_A$ is set at 1.27 [\onlinecite{bib:12}], 
and the phase factor $G^{0\nu}$ is adopted from [\onlinecite{bib:42}], 
$m_e$ is the electron mass.

As shown in Table.~\ref{tab:c1A}, 
compared with our previous 0$\nu\beta\beta$ result 
from a p-type point contact germanium (PPCGe) detector 
in the CDEX-1 experiment [\onlinecite{bib:23}], 
this work achieves a lower background by applying the PSD method. 
The CDEX-1 detector has a lower efficiency mainly due to the 
pulsed-reset preamplifier. The reset preamplifier is designed to have 
low electronic noise for dark matter detection and
has a lower efficiency at the Q$_{\beta\beta}$ energy than the RC preamplifier used in this work.

\begin{table*}[htbp]
 \caption{\label{tab:c1A}Neutrinoless double-beta decay results 
 from this work and CDEX-1. (cpkky denotes counts/kg/keV/yr)}
 \begin{ruledtabular}
   \renewcommand\arraystretch{1.5}
 \begin{tabular}{ccc}
   &BEGe in This work&PPCGe in CDEX-1 \\ \hline
   Exposure & 186.4 kg$\cdot$day & 304 kg$\cdot$day \\
   Total Efficiency & 73.21\% & 68.44\% \\
   \multirow{2}{*}{Background level} & 8.95$\pm$0.22 cpkky (before PSD)
   &\multirow{2}{*}{4.38 cpkky (w/o. PSD)} \\
   &2.35$\pm$0.11 cpkky (after PSD)& \\
   half-life limit (90\% C.L.) & 5.6$\times$10$^{22}$ yr & 6.4$\times$10$^{22}$ yr \\
 \end{tabular}
 \end{ruledtabular}
\end{table*}

\begin{figure}[htb]
\includegraphics
[width=1.0\hsize]
{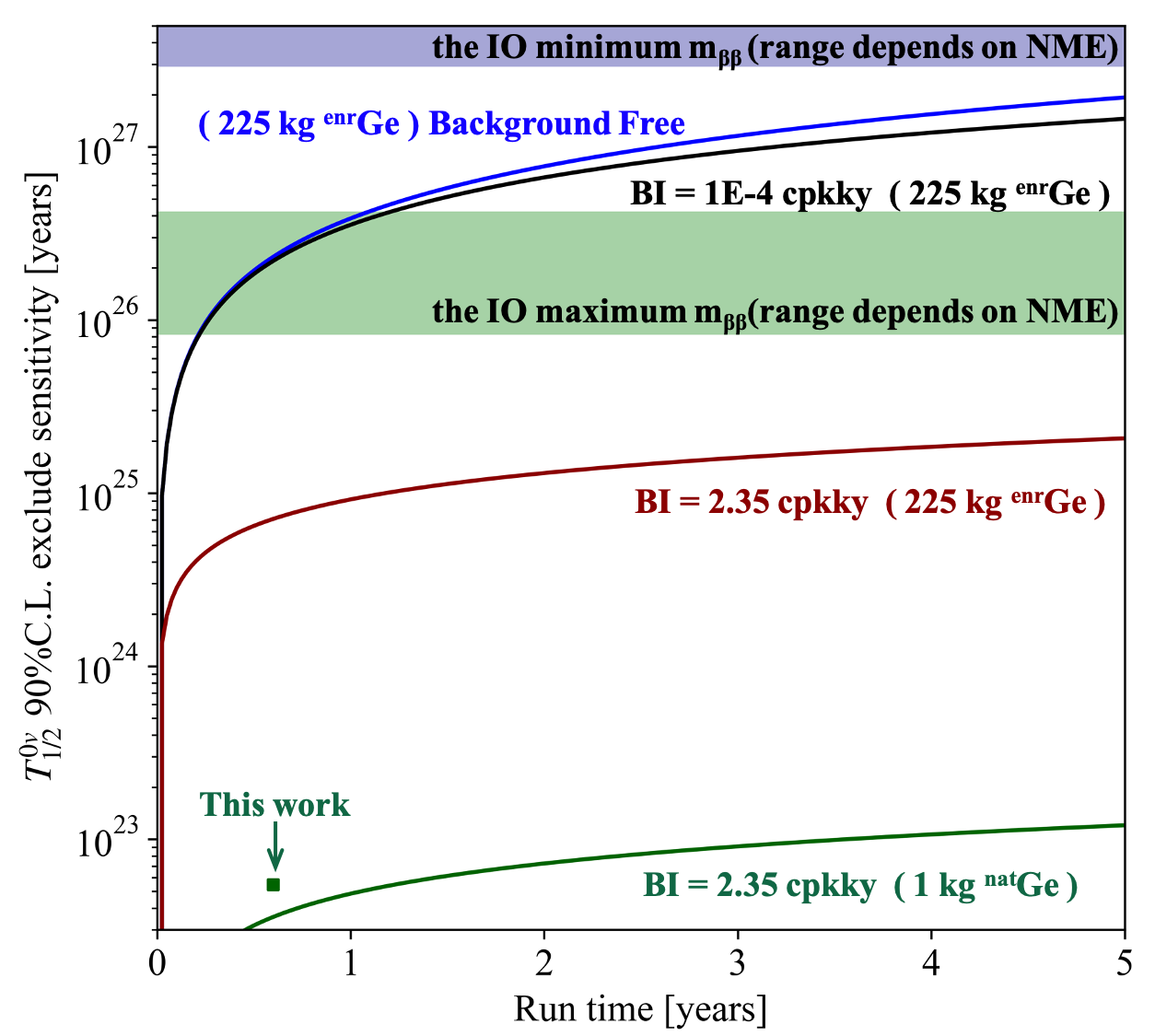}
\caption{\label{fig:sensitivity} The sensitivity 
of 0$\nu\beta\beta$ decay half-life verses operating times 
for different background levels (BI) in the 0$\nu\beta\beta$ signal region, 
$^{76}$Ge enrichments and detector masses. 
The green and blue regions are the 0$\nu\beta\beta$ decay half-life 
corresponding to the upper and lower bound of the
inverted-ordering (IO) neutrino mass scale region[\onlinecite{bib:40}]. 
The range of which is dependent on the uncertainty of 
$^{76}$Ge nuclear matrix element (2.66$\sim$6.04) 
[\onlinecite{bib:40}-\onlinecite{bib:43}]. 
The sensitivities are calculated using the 
approximation outline in [\onlinecite{bib:44}] under the Poisson statistics, 
and the result of this work (the blue square) is calculated 
via an unbinned extended profile likelihood analysis.}
\end{figure}

% As shown in Fig.~\ref{fig:sensitivity}, 
% several key aspects to improve the 0$\nu\beta\beta$ decay 
% half-life sensitivity to meet the 10$^{27}$ yr goal of the 
% next-generation CDEX-300$\nu$ $^{76}$Ge 0$\nu\beta\beta$ experiment 
% can be addressed: 
Future Ge-$0\nu\beta\beta$ experiments would target at 
ton-scale of enriched $^{76}$Ge detectors to probe 
the neutrino inverted mass ordering, 
as is the case for the LEGEND proposal [\onlinecite{bib:43}].
The next-generation CDEX-300$\nu$ experiment will 
consist of 225 kg of Ge-detectors enriched in $^{76}$Ge. 
To meet the half-life sensitivity goal of $10^{27}$ years, 
various improvement will be implemented:
\textit{a:} Increasing the effective exposure: 
The CDEX-300$\nu$ experiment will use 
approximate 225 kg $^{76}$Ge enriched 
($>$86\% enrichment) BEGe detectors to 
increase the exposure.
% search for the 0$\nu\beta\beta$ decay of $^{76}$Ge.
% \paragraph{}

\textit{b:} Background control:
\begin{itemize}
 \item[\textit{b.1}]
 \textit{Deep underground laboratory:} 
 the 2400 m rock overburden of CJPL-II 
 shield the cosmic muons to 
 $O$(10$^{-10}$) cm$^{-2}$s$^{-1}$ [\onlinecite{bib:25}];
 \item[\textit{b.2}] 
 \textit{Cosmogenic background control:}
 measures have been taken to reduce 
 the cosmogenic background.
 The production processes of detectors are 
 optimized to reduce exposures on the ground. 
 Additional shielding is designed to protect the detector 
 from cosmic rays during its production and transportation.
 Continued efforts have been put into  
 underground material production  
 and underground detector fabrication.
\item[\textit{b.3}] 
 \textit{Material screening and selection:}
 materials used in the construction of the next 
 generation experiment will be measured and selected
 according to the physics goal. 
 The detector module will be further optimized to have 
 fewer surrounding structures with higher radiopurity;
 \item[\textit{b.4}]
 \textit{Liquid nitrogen (LN) shielding:}
 the LAr veto system and the detector array will be 
 submerged in a 1725 m$^{3}$ liquid nitrogen (LN) tank. 
 The over 6 m thick LN can provide an effective 
 shield against ambient radioactivity;
 \item[\textit{b.5}]
 \textit{Liquid argon (LAr) veto system:}
 the detector array will be surrounded by 
 LAr coupled with scintillation light readout equipment. 
 Background event causing simultaneous signals 
 in Ge detector and LAr can be rejected
 [\onlinecite{bib:43},\onlinecite{bib:45}];
 \item[\textit{b.6}]
 \textit{Ge detector multiplicity:}
 background events, for instance 
 scattered $\gamma$ rays with simultaneous 
 energy depositions in multiple Ge detectors 
 can be rejected by the coincidence signals
 [\onlinecite{bib:43}];
 \item[\textit{b.7}] 
 \textit{Pulse shape discrimination:}
 for instance,
 A/E method will be used to discriminate MSEs 
 (background-like) from SSEs (signal-like), 
 and the background suppression power of the A/E cut 
 is evaluated to be a factor of 3.79;
%  \item[\textit{b.6}] 
%  \textit{Material screening and selection:}
%  materials used in the construction of the next 
%  generation experiment will be measured and selected
%  according to the physics goal. 
%  The detector module will be further optimized to have 
%  fewer surrounding structures with higher radiopurity;
%  \item[\textit{b.7}] 
%  \textit{Cosmogenic background control:}
%  measures have been taken to reduce 
%  the cosmogenic background.
%  The production processes of detectors are 
%  optimized to reduce exposures on the ground. 
%  Additional shielding is designed to protect the detector 
%  from cosmic rays during its production and transportation.
%  Continued efforts have been put into  
%  underground material production  
%  and underground detector fabrication.
\end{itemize}

Fig.\ref{fig:sensitivity} depicts 
the sensitivity enhancement due to $^{76}$Ge enrichment, 
increased exposure and suppression of background. 
The target sensitivity of CDEX-300$\nu$ is 
$10^{27}$ yr with one ton-yr exposure at a 
background level of $10^{-4}$ cpkkd in the 
$0\nu\beta\beta$ signal region.

\section{\label{sec:5}Summary}

A prototype facility using a natural BEGe detector 
to study the feasibility of building a next generation 
$^{76}$Ge 0$\nu\beta\beta$ experiment
% to search for 0$\nu\beta\beta$ decay in $^{76}$Ge 
is built in this work. 
Event selection and data analysis procedures are established 
to remove unphysical events, reconstruct energy, 
and discriminate background events. 
The pulse shape discrimination method (the A/E cut) is applied 
in data analysis and reduces the background in the 0$\nu\beta\beta$ ROI 
by a factor of 3.79.

A background model is built for the prototype. 
Radionuclides from $^{232}$Th and $^{238}$U decay chains are identified as the 
primary source of backgrounds in the 0$\nu\beta\beta$ signal region. 
Cosmogenic radioactive isotopes in germanium and copper also 
contribute to the backgrounds. 
To control backgrounds in the future large-scale experiment, 
(1) selecting ultra-pure materials can reduce the inhabit radioactive impurities 
from $^{232}$Th and $^{238}$U decay chains, 
(2) growing germanium crystal in an underground facility or 
cooling detector and copper material underground can 
reduce backgrounds from cosmogenic isotopes, 
(3) the anti-coincidence techniques can be used to 
further suppress backgrounds in the 0$\nu\beta\beta$ signal region.
% For the future CDEX-300$\nu$ experiment,
% approaches adopted in its baseline design 
% for background control
Background control approaches adopted in the 
baseline design of the future CDEX-300$\nu$ experiment
are outlined in this work.

Based on the 186.4 kg$\cdot$day exposure data, 
a limit on the half-life of $^{76}$Ge 0$\nu\beta\beta$ decay is set to 
5.62$\times$10$^{22}$ yr at 90\%C.L. 
via an unbinned extended profile likelihood method.

\begin{acknowledgments}
This work was supported by 
the National Key Research and Development Program of China 
(Grant No. 2017YFA0402200) and 
the National Natural Science Foundation of China 
(Grants No. 12175112, No. 12005111, No. 11725522, No. 11675088, No.11475099).
\end{acknowledgments}

\nocite{*}

% \bibliography{PaperRef}% Produces the bibliography via BibTeX.
% \bibliographystyle{plainnat}

\end{document}